\documentclass[aip,twocolumn,reprint,superscriptaddress]{revtex4-1}
\usepackage{amsmath}
\usepackage{amssymb}
\usepackage[latin9]{inputenc}
\usepackage{graphicx}
\usepackage{hyperref}
\usepackage{siunitx}
\usepackage{graphicx}
\usepackage[color= blue!20]{todonotes}
\usepackage[normalem]{ulem}
\usepackage{color}
\usepackage{ulem}
\usepackage[symbol]{footmisc}

\begin{document}

\onecolumngrid

\noindent\textbf{\textsf{\Large Coupling microwave photons to a mechanical resonator using quantum \\ interference}}

\normalsize
\vspace{.3cm}

\noindent\textsf{I.~C.~Rodrigues$^*$, D.~Bothner$^*$, and G.~A.~Steele}

\vspace{.2cm}
\noindent\textit{Kavli Institute of Nanoscience, Delft University of Technology, PO Box 5046, 2600 GA Delft, The Netherlands\\$^*$\normalfont{these authors contributed equally}}

\vspace{.5cm}

\date{\today}

{\addtolength{\leftskip}{10 mm}
\addtolength{\rightskip}{10 mm}

In recent years, the field of microwave optomechanics has emerged as leading platform for achieving quantum control of macroscopic mechanical objects.
Implementations of microwave optomechanics to date have coupled microwave photons to mechanical resonators using a moving capacitance.
While simple and effective, the capacitive scheme suffers from inherent and practical limitations on the maximum achievable coupling strength.
Here, we experimentally implement a fundamentally different approach: flux-mediated optomechanical coupling.
In this scheme, mechanical displacements modulate the flux in a superconducting quantum interference device (SQUID) that forms the inductor of a microwave resonant circuit.
We demonstrate that this flux-mediated coupling can be tuned in-situ by the magnetic flux in the SQUID, enabling nanosecond flux tuning of the optomechanical coupling.
Tuning the external in-plane magnetic transduction field, we observe a linear scaling of the single-photon coupling strength, reaching rates comparable to the current state-of-the-art.
Finally, this linear scaling is predicted to overcome the limits of single-photon coupling rates in capacitive optomechanics, opening the door for a new generation of groundbreaking optomechanical experiments in the single-photon strong coupling regime.

}

\vspace{.5cm}

\twocolumngrid

%\subsection*{Introduction}

%
Parametrically coupling mechanical motion to light fields confined inside a cavity has allowed for major scientific and technological breakthroughs within the recent decade \cite{Aspelmeyer14}.
Such optomechanical systems have been used for sideband-cooling of mechanical motion into the quantum ground state \cite{Teufel11, Chan11}, for the detection of mechanical displacement with an imprecision below the standard quantum limit \cite{Teufel09, Anetsberger10}, for the generation of non-classical mechanical states of motion \cite{Wollman15, Pirkkalainen15, Reed17} and for the entanglement of mechanical oscillators \cite{Riedinger18, OckeloenKorppi18}.
As the mechanical elements can be coupled to both, light fields in the optical and in the microwave domain, current efforts using optomechanical systems target towards the implementation of a quantum link between superconducting microwave quantum processors and optical frequency quantum communication \cite{Bochmann13, Andrews14}.
Another exciting perspective of optomechanical systems is testing quantum collapse and quantum gravity models by preparing Fock and Schroedinger cat states of massive mechanical oscillators \cite{Marshall03, Bahrami14}.
The state transfer fidelity between photons and phonons in optomechanical systems is determined by the coupling rate between the subsystems, and most optomechanical systems so far have single-photon coupling rates much smaller than the decay rates of the cavity.
The strong-coupling regime, necessary for efficient coherent state transfer, is achieved by enhancing the total coupling rate $g=\sqrt{n_c}g_0$ through large intracavity photon numbers $n_c$ \cite{Groeblacher09, Teufel11a, Verhagen12}.
In the optical domain, large photon numbers result in absorption that heats the mechanical mode far above the mode temperature \cite{Meenehan14}. In the microwave domain, large photon numbers result in non-equilibrium cavity noise \cite{Teufel11, Yuan15} that is not completely understood. Both of these sources of noise limit ground state cooling and the fidelity of mechanical quantum ground state preparation.
An approach to reduce these parasitic side-effects is to increase the single-photon coupling rate $g_0$ significantly.
Doing so, optomechanics could even reach the single-photon strong-coupling regime, where the optomechanical system acquires sufficient non-linearity from the parametric coupling such that non-Gaussian mechanical states can be directly prepared by coherently driving the system \cite{Nunnenkamp11, Rabl11}.
In the microwave domain, the most common approach to build an optomechanical system is to combine a superconducting microwave LC circuit with a metallized suspended membrane or nanobeam as mechanical oscillator.
The devices are constructed in a way that the displacement of the mechanical oscillator changes the capacitance of the circuit $C(x)$ and hence its resonance frequency $\omega_0(x) = 1/\sqrt{LC(x)}$.
In this configuration, however, the single-photon coupling rate is limited to $g_0 \leq \frac{\omega_0}{2}\frac{x_\mathrm{zpf}}{d}$ with the zero-point fluctuation amplitude $x_\mathrm{zpf}$ and the capacitor gap $d$.
Current devices are highly optimized, but still achieve typically only $x_\mathrm{zpf}/d \approx 10^{-7}$ for a parallel plate capacitor gap of $d=50\,$nm and it is extremely challenging to increase $g_0$ beyond $300 \,$Hz with this approach.
\begin{figure*}
	\centerline{\includegraphics[trim = {0.2cm, 1.7cm, 1.2cm, 0.5cm}, clip=True, width=0.97\textwidth]{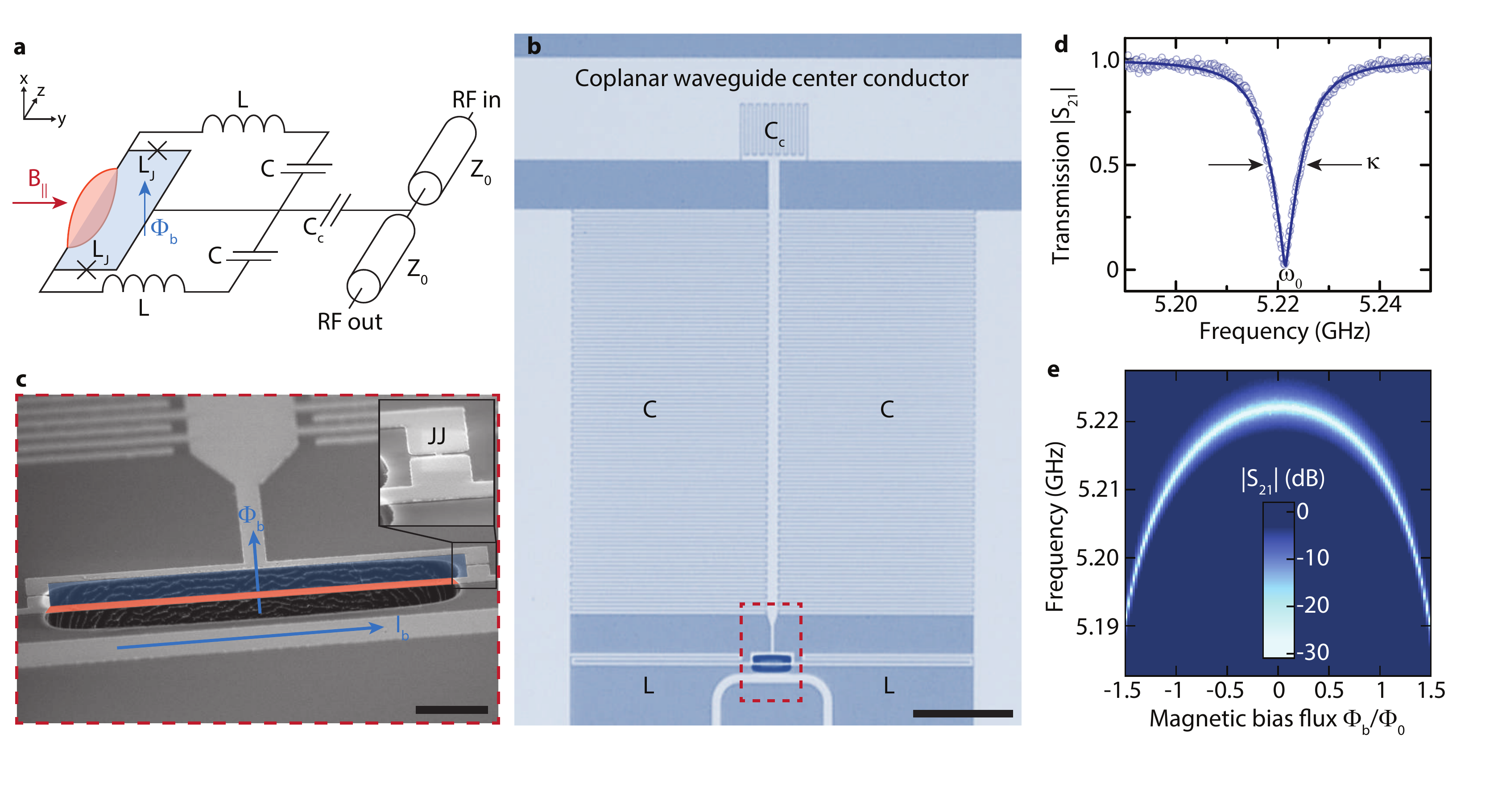}}
	\caption{\textsf{\textbf{A superconducting microwave circuit with magnetic-flux mediated optomechanical coupling to a mechanical oscillator.} \textbf{a} Circuit schematic of the device. The LC circuit is capacitively coupled to a microwave transmission line with characteristic impedance $Z_0$ by means of a coupling capacitor $C_c$. In addition to the linear capacitors $C$ and inductors $L$, a superconducting quantum interference device (SQUID) is built into the circuit, consisting of two Josephson junctions with inductance $L_J$ in a closed superconducting loop, of which a part is suspended and free to move perpendicular to the circuit plane. To bias the SQUID with magnetic flux $\Phi_b$, a magnetic field can be applied perpendicular to the circuit plane. Motion of the mechanical element is transduced into modulations of the bias flux by a magnetic in-plane field $B_{||}$. An optical micrograph of the circuit is shown in \textbf{b}, light gray parts correspond to a $20\,$nm thick layer of aluminum, dark parts to silicon substrate. The black scale bar corresponds to $50\,\mu$m. The red dashed box shows the region, which is depicted in a tilted scanning electron micrograph in \textbf{c}, showing the SQUID loop with the released aluminum beam. The bias flux through the SQUID loop $\Phi_b$ can be changed by a bias current $I_b$ sent through the on-chip flux bias line. The black scale bar corresponds to $3\,\mu$m. The inset shows a zoom into one of the constriction type Josephson junctions (JJs). In \textbf{d} the cavity resonance is shown, measured by sending a microwave tone to the microwave feedline and detecting the transmitted signal $S_{21}$. A fit to the data points (circles), shown as line, reveals a resonance frequency of $\omega_0 = 2\pi\cdot5.221\,$GHz and a linewidth $\kappa = 2\pi\cdot9\,$MHz. Panel \textbf{e} shows color-coded the tuning of the cavity resonance absorption dip with magnetic bias flux in units of flux quanta $\Phi_b/\Phi_0$.}}
	\label{fig:Device}
\end{figure*}
Here, we realize a fundamentally different approach for a microwave optomechanical device by incorporating a suspended mechanical beam into the loop of a superconducting quantum interference device (SQUID).
The SQUID itself is part of a superconducting LC circuit and essentially acts as an inductor, whose inductance depends on the magnetic flux threading through the loop.
In contrast to the capacitive approach, this magnetic flux-mediated inductive coupling scheme provides quickly tunable single-photon coupling rates \cite{Via15, Nation16}, which in addition scale linearly with a magnetic field applied in the plane of the SQUID loop \cite{Shevchuk17}.
In contrast to capacitive microwave optomechanics, the coupling rates are not limited by geometric and technological restrictions and there is a realistic prospective for achieving the optomechanical single-photon strong coupling regime.
%

%
%\section*{Results}
%

%
\begin{figure*}
	\centerline{\includegraphics[clip=True, width=0.97\textwidth]{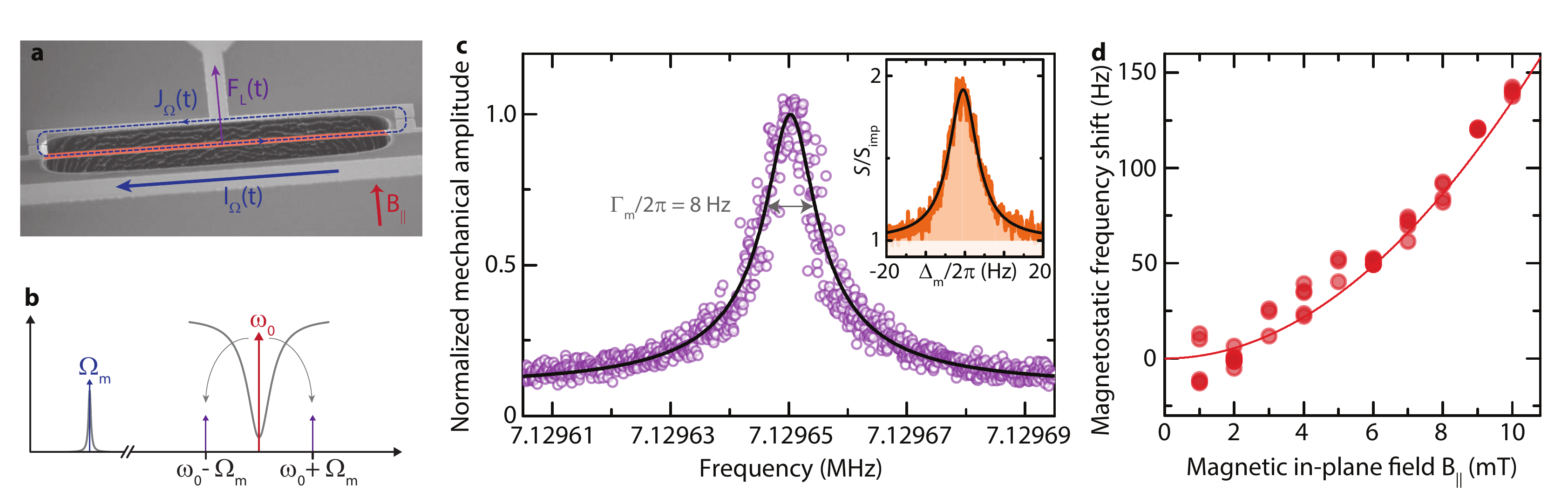}}
	\caption{\textsf{\textbf{Detection of mechanical motion using a superconducting SQUID cavity interferometer and observation of magnetostatic spring stiffening.} \textbf{a} Schematic of coherently driving the mechanical oscillator by means of the Lorentz force. The current sent through the bias line has a DC component to bias the SQUID with a flux $\Phi_b$. This generates a circulating current $J$ in the SQUID loop. In addition, an oscillating current is sent through the line with a frequency close to the mechanical oscillator resonance frequency $\Omega \approx \Omega_m$. Thus, the loop current through the mechanical beam oscillates correspondingly, leading to an oscillating Lorentz force $F_\mathrm{L}(t)$ due to the presence of the magnetic in-plane field $B_{||}$. The mechanical motion modulates the total magnetic flux through the SQUID loop and hence the cavity resonance frequency. When a resonant coherent microwave tone is sent into the cavity, the mechanical oscillations generate sidebands at $\omega = \omega_0 \pm \Omega$, cf. panel \textbf{b}, which are observed to detect the mechanical motion. In \textbf{c} the down-converted sideband signal is shown during a sweep of the excitation frequency $\Omega$. Circles are data, the line is a Lorentzian fit and both are normalized to the maximum of the fit curve. The inset depicts the down-converted sideband thermal noise spectral density in absence of a coherent driving force, normalized to the background noise floor. Orange line are data, black line is a Lorentzian fit. The contribution from the background noise is shaded in white and the contribution from the mechanical displacement noise in orange. When increasing the magnetic in-plane field, we observe a shift of the mechanical oscillator resonance frequency, shown in panel \textbf{d}. This frequency shift is induced by a position-dependent contribution to the Lorentz-force and corresponds to a magnetostatic stiffening of the mechanical spring constant. The circles are data and the line corresponds to a theoretical curve with $\delta\Omega_m\propto B_{||}^2$.}}
	\label{fig:Mechanics}
\end{figure*}
%

%
%\subsection*{Results: Concept and device}
%

%
The concept of coupling mechanical resonators to SQUIDs has been developed in many works \cite{Xue07, Wang08}, including earlier experimental work with DC SQUIDs \cite{Etaki08, Poot10}.
Recently, this concept was extended theoretically to optomechanics \cite{Shevchuk17}, describing a way using SQUIDs to achieve strong and tunable optomechanical coupling between a vibrating beam and a superconducting cavity.
The circuit used here for its realisation is schematically shown in Fig.~\ref{fig:Device}\textbf{a}.
The idea is based on transducing mechanical displacement to magnetic flux, which in turn modulates the effective inductance of a SQUID and therefore the resonance frequency of the LC circuit hosting it.
To achieve this transduction from displacement to flux, a part of the SQUID loop is suspended and the device is exposed to an external magnetic field $B_{||}$ applied parallel to the device plane.
The suspended loop part acts as a mechanical beam resonator and its vibrational motion, perpedicular to the device plane, will create an effective SQUID area perpendicular to the applied field $B_{||}$, i.e., couple a net magnetic flux into the loop.
The inductance $L(\Phi_b)$ of an LC circuit containing a SQUID depends on the magnetic flux threading the SQUID loop, and translates to a flux-dependent resonance frequency
\begin{equation}
\omega_0(\Phi_b) = \frac{1}{\sqrt{L(\Phi_b)C}}.
\end{equation}
When the displacement of a mechanical oscillator is transduced to additional flux, an optomechanical interaction between mechanical mode and cavity resonance frequency emerges and the single-photon coupling rate is given by \cite{Shevchuk17}
\begin{equation}
g_0 = \frac{\partial\omega_0}{\partial\Phi}\Phi_\mathrm{zpf} = \frac{\partial\omega_0}{\partial\Phi}\gamma B_{||}lx_\mathrm{zpf}.
\end{equation}
The first term $\partial\omega_0/\partial\Phi$ corresponds to the responsivity of the SQUID cavity resonance frequency to small changes of flux through the loop and allows for very fast tuning of $g_0$.
The second term $\Phi_\mathrm{zpf} = \gamma B_{||}l x_\mathrm{zpf}$ is the magnetic flux fluctuation induced in the SQUID by the mechanical zero-point fluctuations $x_\mathrm{zpf}$ of the beam with length $l$ and scales linearly with an in-plane magnetic field $B_{||}$, cf. Fig.~\ref{fig:Device}.
The scaling factor $\gamma$ accounts for the mode shape of the mechanical oscillations and is on the order of 1.
The microwave SQUID cavity in our experiment is made of a single $20\,$nm thick layer of sputtered aluminum on a silicon substrate and it contains a SQUID consisting of two constriction-type Josephson junctions placed in parallel in a $21 \times 5\,\mu$m$^2$ closed loop.
An optical image of the device is shown in Fig.~\ref{fig:Device}\textbf{b} and an electron microscope image of the SQUID loop in \textbf{c}, the fabrication is detailed in the Supplementary Material Sec.~S1.
The capacitance of the LC circuit is formed by two interdigitated capacitors $C$ to ground and a coupling capacitor $C_c$ to the center conductor of a coplanar waveguide feedline.
Additionally to the SQUID inductance $L_S = L_J/2$, there are two linear inductances $L$ built into the circuit in order to dilute the non-linearity of the cavity, arising from the non-linear Josephson inductance.
By this measure we achieve an anharmonicity of approximately $15\,$Hz per photon and enable the multi-photon coupling rate enhancement $g=\sqrt{n_c}g_0$ of linearized optomechanics.
The cavity is side-coupled to a coplanar waveguide microwave feedline, which is used to drive and read-out the cavity response by means of the transmission parameter $S_{21}$.
The device is mounted into a radiation tight metal housing and attached to the mK plate of a dilution refrigerator with a base temperature of approximately $T_b = 15\,$mK, cf. SM Sec.~S2.
Without any flux biasing, the cavity has a resonance frequency $\omega_0 = 2\pi\cdot5.221\,$GHz and a linewidth $\kappa = 2\pi\cdot9\,$MHz, which at the same time corresponds to the external linewidth $\kappa\approx\kappa_e$ due to being deep in the over-coupled regime, cf. the cavity resonance curve shown in Fig.~\ref{fig:Device}\textbf{d}.
When magnetic flux is applied to the SQUID loop by sending a current to the chip via the on-chip flux bias line, the cavity resonance frequency is shifted towards lower values due to an increase of the Josephson inductances inside the SQUID.
The flux-dependent transmission $|S_{21}|(\Phi)$ is shown in Fig.~\ref{fig:Device}\textbf{e} and a total tuning of about $30\,$MHz can be achieved, mainly limited by a non-negligible SQUID loop-inductance of the SQUID and the dilution of the Josephson inductance by $L_J/(L+L_J) \approx 0.01$, see also SM Sec.~S3.
The largest flux responsivities we could achieve here were approximately $\partial\omega_0/\partial\Phi = 70\,$MHz$/\Phi_0$.
The mechanical oscillator is a $20 \times 1\,\mu$m$^2$ large aluminum beam and is suspended as result of releasing part of the superconducting loop forming the SQUID by removing the underlying silicon substrate in an isotropic reactive ion etching process \cite{Norte18}.
The beam has a total mass $m = 1\,$pg and its fundamental out-of-plane mode oscillates at a frequency $\Omega_m = 2 \pi \cdot 7.129\,$MHz with an intrinsic mechanical damping rate of $\Gamma_m = 2\pi \cdot 8\,$Hz or quality factor $Q_m = \Omega_m/\Gamma_m \approx 9\cdot10^{5}$, which is exceptionally high for a mechanical oscillator made from a sputter-deposited metal film. 
From the mass and resonance frequency, the zero-point motion of the oscillator is estimated to be $x_\mathrm{zpf} = \sqrt{\frac{\hbar}{2m\Omega_m}} = 33\,$fm.
%

%\subsection*{Results: Interferometric characterization of mechanical oscillator}

%
The mechanical beam can be coherently driven by Lorentz-force actuation using the on-chip flux bias line.
When a current is sent through the bias line, magnetic flux is coupled into the SQUID loop and a circulating loop current is flowing through the mechanical oscillator.
We apply a current $I_\Omega(t) = I_\mathrm{dc} + I_0\cos{\Omega t }$ with $\Omega \approx \Omega_m$, where the DC component $I_\mathrm{dc}$ is simultaneously biasing the SQUID and -- in presence of an in-plane magnetic field $B_{||}$ -- exerting a constant Lorentz force to the beam. 
The oscillating part $I_0\cos{\Omega t}$ modulates the total Lorentz-force $F_\mathrm{L}(t) = F_\mathrm{dc}+F_0\cos{\Omega t}$ around the equilibrium value $F_\mathrm{dc}$ and effectively drives the mechanical oscillator.
The concept is illustrated in Fig.~\ref{fig:Mechanics}\textbf{a}, for more details cf. SM Sec.~S4. 
\begin{figure}
	\centerline{\includegraphics[trim = {0cm, 0.3cm, 0cm, 0cm}, clip=True, scale=.5]{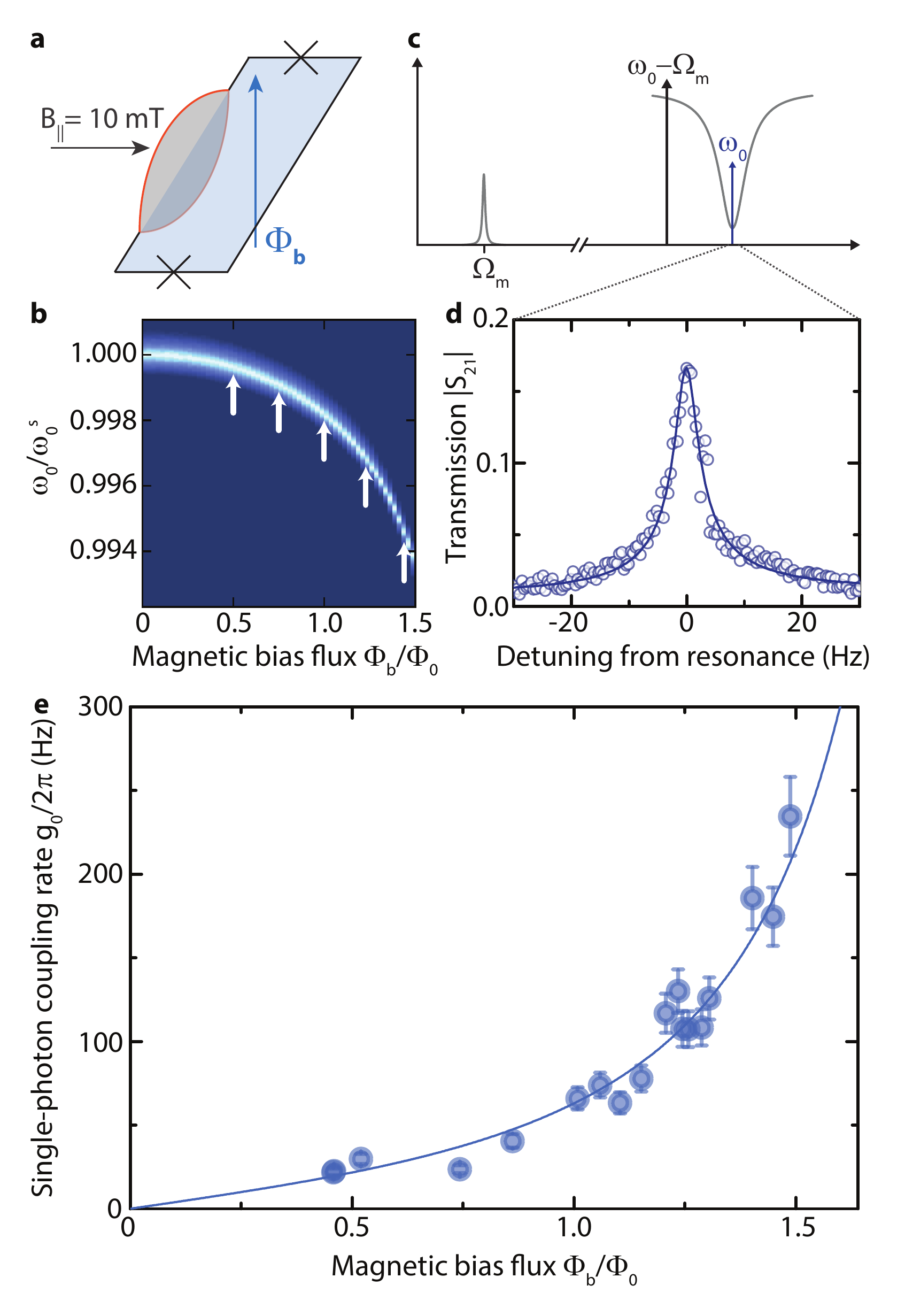}}
	\caption{\textsf{\textbf{Tuning the optomechanical single-photon coupling rate by changing the flux operating point of the SQUID}} \textbf{a} Schematic of the applied magnetic field components to the SQUID loop. The in-plane magnetic field $B_{||}$ is set by means of a cylindrical coil wrapped around the whole sample mounting. During this experiment, it was kept constant at $B_{||} = 10\,$mT. Additionally, an out-of-plane magnetic field was varied by changing the current sent through the on-chip flux bias line, generating a magnetic bias flux $\Phi_b$. \textbf{b} As consequence of changing the amount of flux threading the SQUID loop, both the resonance frequency as well as the flux responsivity $\partial\omega_0/\partial\Phi$ of the cavity are changed. The plot shows $|S_{21}|$ ($B_{||} = 1\,$mT), the color code is given in Fig.\ref{fig:Device}e. The white arrows represent the points, for which we performed the measurement scheme of optomechanically induced transparency (OMIT) as shown schematically in \textbf{c}. A coherent drive tone is set to the red sideband of the SQUID cavity $(\omega_d = \omega_0 - \Omega_m)$, while a small probe tone is scanning the cavity resonance $\omega_p \approx \omega_0$. As result of an interference effect, a transparency window in the transmitted signal $S_{21}$ is visible around $\omega_d + \Omega_m$, as shown in \textbf{d}, where the circles represent the data and the line the corresponding fit curve. By setting the cavity to different flux bias points (white arrows in \textbf{b}), we change the cavity flux responsivity and therefore the single-photon optomechanical coupling rate $g_0 \propto \partial\omega_0/\partial\Phi$. From the magnitude of the transparency window, $g_0$ can be extracted for each flux bias point. The result is plotted in \textbf{e} as circles. The line is the theoretical curve as described in the main text.}
	\label{fig:OMIT}
\end{figure}
The resulting mechanical motion modulates the cavity resonance frequency and generates sidebands at $\omega_d \pm \Omega$ to a microwave signal sent into the cavity at $\omega_d = \omega_0$, cf. the schematic in Fig.~\ref{fig:Mechanics}\textbf{b}.
By sweeping $\Omega$ through $\Omega_m$ and down-converting the sidebands generated at $\omega = \omega_0 - \Omega$ and $\omega = \omega_0 + \Omega$, we measure the mechanical resonance as shown in Fig.~\ref{fig:Mechanics}\textbf{c}.
This interferometric detection scheme of displacement can also be used to detect the thermal motion of the mechanical oscillator.
At the dilution refrigerator base temperature $T_b = 15\,$mK, we expect a thermal mode occupation of the beam of approximately $n_\mathrm{th} = k_\mathrm{B}T_b/\hbar\Omega_m \approx 46$ phonons with $k_\mathrm{B}$ being the Boltzmann constant.
In the inset of Fig.~\ref{fig:Mechanics}\textbf{c} we show the down-converted sideband power spectral density $S$ of the cavity output field, normalized to the background noise, without any external drive applied to the mechanical oscillator.
On top of the imprecision noise background $S_\mathrm{imp}$ of the measurement chain, a Lorentzian peak with a linewidth of $\sim 8\,$Hz is visible, generated by the residual thermal motion of the beam.
When we sweep the magnetic in-plane field $B_{||}$, we observe an increase of the mechanical resonance frequency as shown in Fig.~\ref{fig:Mechanics}\textbf{d} induced by Lorentz-force backaction \cite{Poot10}.
Complementary to the electrostatic spring softening in mechanical capacitors with a bias voltage, this effect can be understood as a magnetostatic spring stiffening.
When the mechanical oscillator is displaced from its equilibrium position, an additional magnetic flux is coupled into the SQUID loop, which leads to an adjustment of the circulating current $J$ to fulfill fluxoid quantization inside the loop.
Hence, the Lorentz-force $F_\mathrm{L} \propto B_{||}J$ will change accordingly and therefore has a contribution dependent on the mechanical position.
For small mechanical amplitudes and circulating currents not too close to the critical current of the Josephson junctions, this position dependence will be linear, causing a frequency shift $\delta\Omega_m \propto B_{||}^2$, cf. the discussion in the Supplementary Material Sec.~S4.
%

%\subsection*{Tuning the optomechanical single-photon coupling rate}

%
When a magnetic bias flux is applied to the SQUID, not only the cavity resonance frequency changes, but also the flux responsivity $\partial\omega_0/\partial\Phi$.
As the optomechanical single-photon coupling rate is directly proportional to the responsivity, it can in principle be switched on and off on extremely short timescales or can be dynamically controlled by flux modulating the SQUID.
We demonstrate this tuning of the single-photon coupling rate with bias flux by determining $g_0$ for different values of $\Phi_b/\Phi_0$.
\begin{figure}
	\centerline{\includegraphics[trim = {0.0cm, 6cm, 0.0cm, 0.0cm}, clip=True, scale=.5]{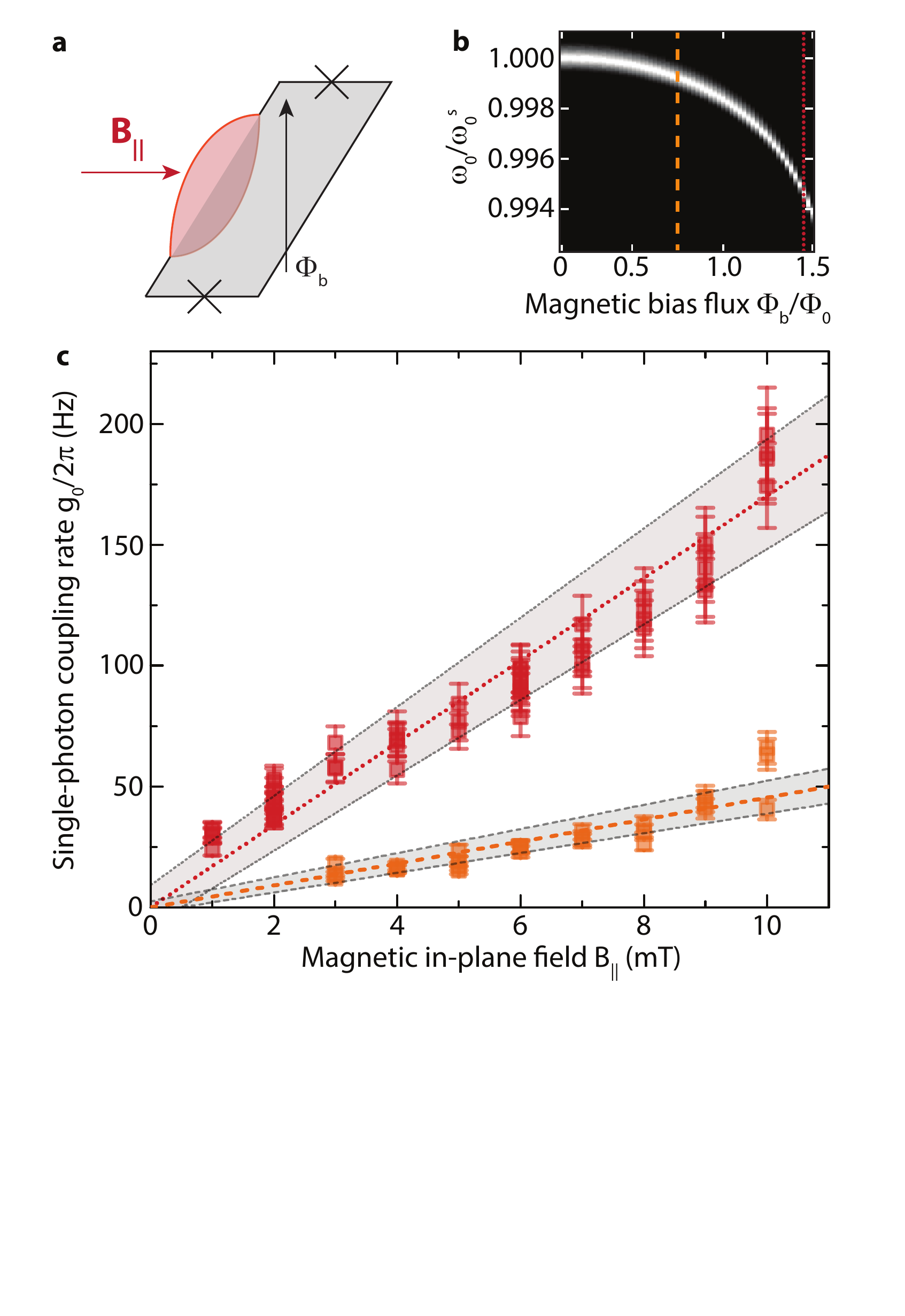}}
	\caption{\textsf{\textbf{Scaling up the optomechanical single-photon coupling rate with the applied in-plane magnetic field.} \textbf{a} Representation of the applied magnetic field components to the SQUID loop. During the experiment, the cavity flux responsivity was fixed at two different values by adjusting the flux bias point $\Phi_b$. In addition to this constant parameter, the in-plane magnetic field $B_{||}$ was swept from $1$ to $10\,$mT in steps of $1\,$mT. The transmission $|S_{21}|$ depending on the normalized bias flux is shown in \textbf{b} for $B_{||}=1\,$mT (black: $0\,$dB, white: $-30\,$dB). The two different set-points represented as orange dashed and red dotted lines, respectively, correspond to a flux responsivity of $\sim 17\,$MHz$/\Phi_0$ and $\sim 60\,$MHz$/\Phi_0$. Posterior to tuning the cavity to the desired working point, an OMIT experiment was performed and the single-photon coupling rate of the system was extracted. The experimental procedure was repeated in increasing steps of $1\,$mT of in-plane field. The resulting single-photon coupling rates $g_0$ are shown in \textbf{c} as squares. The dashed and dotted lines show theoretical lines and the gray areas consider uncertainties in the flux responsivity of $10\%$ and a possible in-plane field offset of $\pm0.5\,$mT.}}
	\label{fig:InPlaneB}
\end{figure}
One possibility to determine the multi-photon coupling rate $g$ in an optomechanical system is to perform the experimental scheme of optomechanically induced transparency \cite{Agarwal10, Weis10}.
For this scheme, a strong coherent microwave tone is driving the cavity on the red sideband $\omega_d = \omega_0 - \Omega_m$ and a weak probe tone is sent to the cavity around $\omega_p \approx \omega_0$.
The two tones interfere inside the cavity, resulting in an amplitude beating with the frequency difference $\Omega = \omega_p - \omega_d$.
If the beating frequency is resonant with the mechanical mode, the radiation pressure force resonantly drives mechanical motion which, in turn, modulates the cavity resonance and the red sideband drive tone.
The modulation generates a sideband to the drive at $\omega = \omega_d + \Omega$, which interferes with the original probe field in the cavity.
This interference effect opens up a narrow transparency window within the cavity response, which has the shape of the mechanical resonance, modified by the dynamical backaction of the red sideband tone.
For $\omega_d = \omega_0 - \Omega_m$ the magnitude of the transparency window $|S_m|$ with respect to the depth of the cavity resonance dip $|S_c|$ is directly related to the coupling rate via
\begin{equation}
\frac{|S_m|}{|S_c|} = \frac{4g^2}{\kappa\Gamma_\mathrm{eff}}
\end{equation}
where $\Gamma_\mathrm{eff} = \Gamma_m + \Gamma_o$ is the width of the transparency window, given by the intrinsic mechanical damping $\Gamma_m$ and the optomechanically induced damping $\Gamma_o$.
In combination with a careful calibration of the intracavity photon numbers $n_c$, we use this approach to get an estimate for the single-photon coupling rate $g_0 = g/\sqrt{n_c}$.
More details on the photon number calibration and the extraction of $g$ from the OMIT data are given in the Supplemetary Material Sec.~S5.
When performing this experiment for several different flux bias points, we find a clear increase of $g_0$ with the cavity flux responsivity.
The experimental scheme and the obtained single-photon coupling rates for a constant in-plane field of $B_{||} = 10\,$mT are shown in Fig.~\ref{fig:OMIT}.
In Fig.~\ref{fig:OMIT}\textbf{e} we also plot as line the theoretical curve, where the only free parameter is the scaling factor $\gamma = 0.86$, taking into account the mode shape of the mechanical oscillations.
All other contributions to the calculations were obtained from independent measurements, such as the bias flux dependence of the cavity frequency, the mechanical resonance frequency and estimations for the beam length and its mass.
The largest single-photon coupling rate we achieve here $g_0 \approx 2\pi\cdot230\,$Hz is comparable with the best values obtained for highly optimized capacitively coupled devices. 
As it is possible to achieve responsivities of several GHz$/\Phi_0$ with SQUID cavities, we expect that with an optimized cavity it is possible to boost the single-photon coupling rates to the order of $\sim 10\,$kHz per mT of in-plane field.
From the Kerr-nonlinearity of our device $\chi/2\pi \sim 120\,$Hz for the largest measured responsivity, we estimate intracavity photon numbers up to $\sim 10^5$ to be compatible with the cavity, which corresponds to maximally achievable multi-photon coupling rates of $g = 2\pi \cdot 70\,$kHz and cooperativities of $C \sim 300$. 
Due to the large loop-inductance of the used SQUID, however, the cavity is operated in a metastable flux branch (see SM Sec.~S3) and we were limited to work with $n_c \sim 150$ intracavity photons before switching to the stable flux branch, which limited $g$ and $C$ to $g \sim 2\pi\cdot 3\,$kHz and $C = 0.5$ in current experiments.

As an ultimate experimental signature that our device transduces mechanical displacement to magnetic flux, we investigate the scaling of the optomechanical coupling rate with magnetic in-plane field $B_{||}$.
Therefore, we performed the scheme of optomechanically induced transparency for constant values of flux responsivity $\partial\omega_0/\partial\Phi$ but for varying in-plane magnetic field.
First, we chose a fixed flux biasing point of about $\Phi_b/\Phi_0 \approx 0.75$ and then adjusted $B_{||}$ in steps of $1\,$mT.
For each $B_{||}$ we perform several OMIT experiments and extract the single-photon coupling rates as described above.
This whole scheme was repeated for $\Phi_b/\Phi_0 \approx 1.45$.
The resulting single-photon coupling rates are shown in Fig.~\ref{fig:InPlaneB} and follow approximately a linear increase with in-plane magnetic field.
The theoretical lines correspond to independent calculations based on the flux-dependence of the cavity, and the parameters of the mechanical oscillator.
The data clearly demonstrate that we observe a flux-mediated optomechanical coupling, a system in which the coupling rates can be further increased with higher magnetic in-plane fields.
In the current setup, we were limited to the field range up to $10\,$mT.
Due to an imperfect alignment between the chip and the in-plane field, a considerable out-of-plane component was present and, most probably by introducing vortices, strongly influenced the properties of the cavities above $B_{||} = 10\,$mT.
Using a vector magnet to compensate for possible misalignments will allow to go up to about $100\,$mT with thin film Aluminum devices \cite{Meservey71, Antler13} resulting in rate of $g_0 \approx \,$MHz.
When extending our materials to other superconductors such as Niobium or Niobium alloys, where similar constriction type SQUIDs have recently been used for tunable resonators \cite{Kennedy19}, the possible field range for the in-plane field increases up to the Tesla regime.
With the realisation of flux-mediated optomechanical coupling reported in this article, the door is opened for a new generation of microwave optomechanical systems.
The single-photon coupling rates achieved with this first device are already competing with the best electromechanical systems and can be boosted towards the MHz regime by optimizing flux responsivity and applying higher magnetic in-plane fields.
In addition, reducing the cavity linewidth to values of $\sim\,100\,$kHz will lead us directly into the single-photon strong-coupling regime, where a new type of devices and experiments can be realized, amongst others the realization of a new class of microwave qubits, where the nonlinearity arises from the coupling to a mechanical element, the generation of mechanical quantum states or optomechanically induced photon blockade.
The coupling mechanism between a mechanical oscillator and a microwave circuit, which we realised here, has also been intensely discussed in the context of superconducting flux and transmon qubits instead of linear cavities \cite{Xue07, Wang08, Wang18, Kounalakis19} and could now be implemented using circuits with a large Josephson non-linearity leading to a new regime of quantum control of macroscopic mechanical objects.

\subsection*{Acknowledgements}
\vspace{-2mm}

The authors thank R. Norte for help with the device fabrication and M.~D. Jenkins for support with the data acquisition software.
This research was supported by the Netherlands Organisation for Scientific Research (NWO) in the Innovational Research Incentives Scheme -- VIDI, project 680-47-526, the European Research Council (ERC) under the European Union's Horizon 2020 research and innovation programme (grant agreement No 681476 - QOMD) and from the European Union's Horizon 2020 research and innovation programme under grant agreement No 732894 - HOT.

\subsection*{Author contributions}
\vspace{-2mm}

I.~C.~R. and D.~B. designed and fabricated the device, performed the measurements and analysed the data.
G.~A.~S. conceived the experiment and supervised the project.
All authors wrote the manuscript and all authors discussed the results and the manuscript.

\subsection*{Competing financial interest}
\vspace{-2mm}
The authors declare no competing financial interests.

\subsection*{Data availability statement}
\vspace{-2mm}

Raw data and processing scripts will be made publically available on zenodo upon acceptance of the manuscript.

\newpage

\widetext

\noindent\textbf{\textsf{\Large Supplementary Material: Coupling microwave photons to a mechanical resonator using quantum interference}}

\normalsize
\vspace{.3cm}

\noindent\textsf{I.~C.~Rodrigues$^*$, D.~Bothner$^*$, and G.~A.~Steele}

\vspace{.2cm}
\noindent\textit{Kavli Institute of Nanoscience, Delft University of Technology, PO Box 5046, 2600 GA Delft, The Netherlands\\$^*$\normalfont{these authors contributed equally}}

\renewcommand{\thefigure}{S\arabic{figure}}
\renewcommand{\theequation}{S\arabic{equation}}

\renewcommand{\thesection}{S\arabic{section}}
\renewcommand{\bibnumfmt}[1]{[S#1]}

\setcounter{figure}{0}
\setcounter{equation}{0}

	\section{Device fabrication}

	\begin{figure}[h]
		\centerline {\includegraphics[trim={0cm 2cm 0cm 2.3cm},clip=True,scale=0.66]{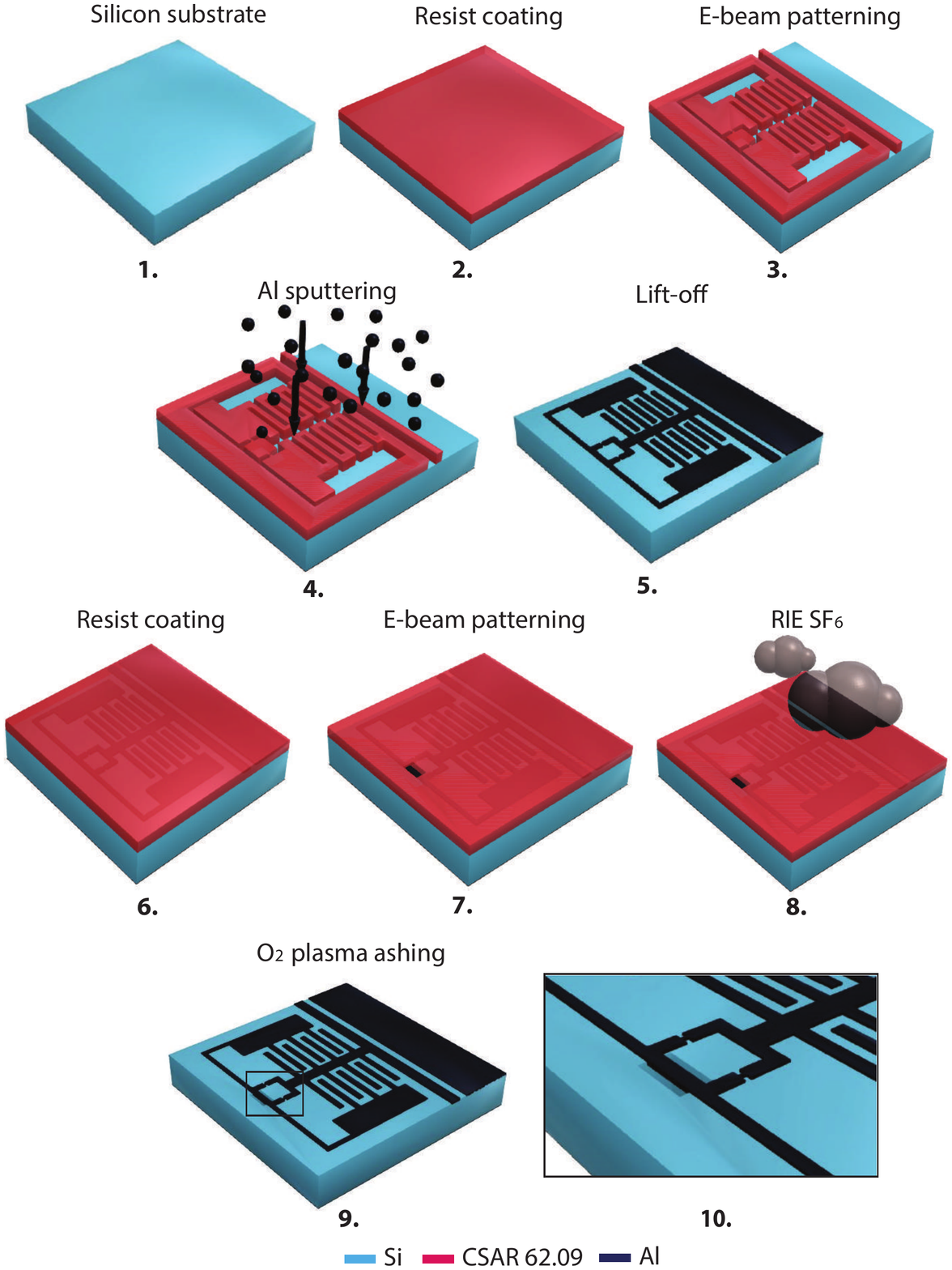}}
		\caption{\textsf{\textbf{Schematic device fabrication.} \textbf{a} \textbf{1.-5.} show the deposition and patterning of the superconducting microwave structures and \textbf{6.-9.} show the etching window patterning and nanobeam release. \textbf{10.} Zoom-in of a SQUID loop with a released beam. Dimensions are not to scale. A description of the individual steps is given in the text.}}
		\label{fig:Fab}
	\end{figure}
	\newpage
	The fabrication of the device starts by patterning alignment markers (made of a $100\,$nm layer of sputter-deposited Molybdenum-Rhenium alloy) on top of a 2 inch silicon wafer.
	CSAR62.13 was used as patterning mask for the subsequent EBL (Electron Beam Lithography) step and warm Anisole at $\sim80\,^{\circ}{\rm C}$ as solution for the lift-off process. 
	Afterwards the wafer was diced in  $14\times14\,$mm$^2$ chips which were then individually used for the following fabrication steps.
	The superconducting structures were patterned in a single EBL step where CSAR62.09 was used as resist.
	Posterior to the exposure, the sample was developed in Pentylacetate for $60\,$seconds followed by a solution of MIBK:IPA (1:1) for another $60\,$seconds and finally rinsed in IPA. 
	Once the mask was developed, the chip was loaded into a sputtering machine where a thin $20\,$nm layer of Aluminum ($1\%$ Silicon) was deposited after a short in-situ cleaning step by means of Argon ion milling.
	After the deposition, the sample was placed in the bottom of a beaker containing a small amount of room-temperature Anisole and left in a ultrasonic bath for a few minutes.
	During this time, the patterning resist is dissolved and the Aluminum layer sitting on top is lifted off. 
	At this point in the fabrication all the superconducting structures were patterned, leaving the most sensitive step, the mechanical release, for the end.
	Before the final release, however, the sample is once again diced to a smaller $10\times10\,$mm$^2$ size in order to fit into the sample mountings and PCBs (Printed Circuit Boards).
	For the final EBL step, a CSAR62.09 resist was once again used as mask and the development of the pattern was done in a similar way as for the first layer.
	Once the etch mask (consisting of two small windows enclosing one arm of the SQUID loop) was patterned, the sample underwent an isotropic SF$_6$ etch (at approx. ($\sim-10\,^{\circ}{\rm C}$) for two minutes.
	During this time the substrate under the beam is etched without attacking the thin aluminum layer forming the cavity and the mechanical beam.
	Once the beams are released, we proceeded with a O$_2$ plasma ashing step in order to remove the remaining resist from the sample.
	In the end of the fabrication, the sample is glued to a PCB and wirebonded both to ground and to the $50\,\Omega$ connector lines. 
	A schematic representation of this fabrication process can be seen in Fig.~\ref{fig:Fab}, omitting the patterning of the electron beam markers.

	\section{Measurement setup}

	\begin{figure}
		\centerline{\includegraphics[trim={0cm 0cm 1cm 0.5cm},clip=True,scale=0.81]{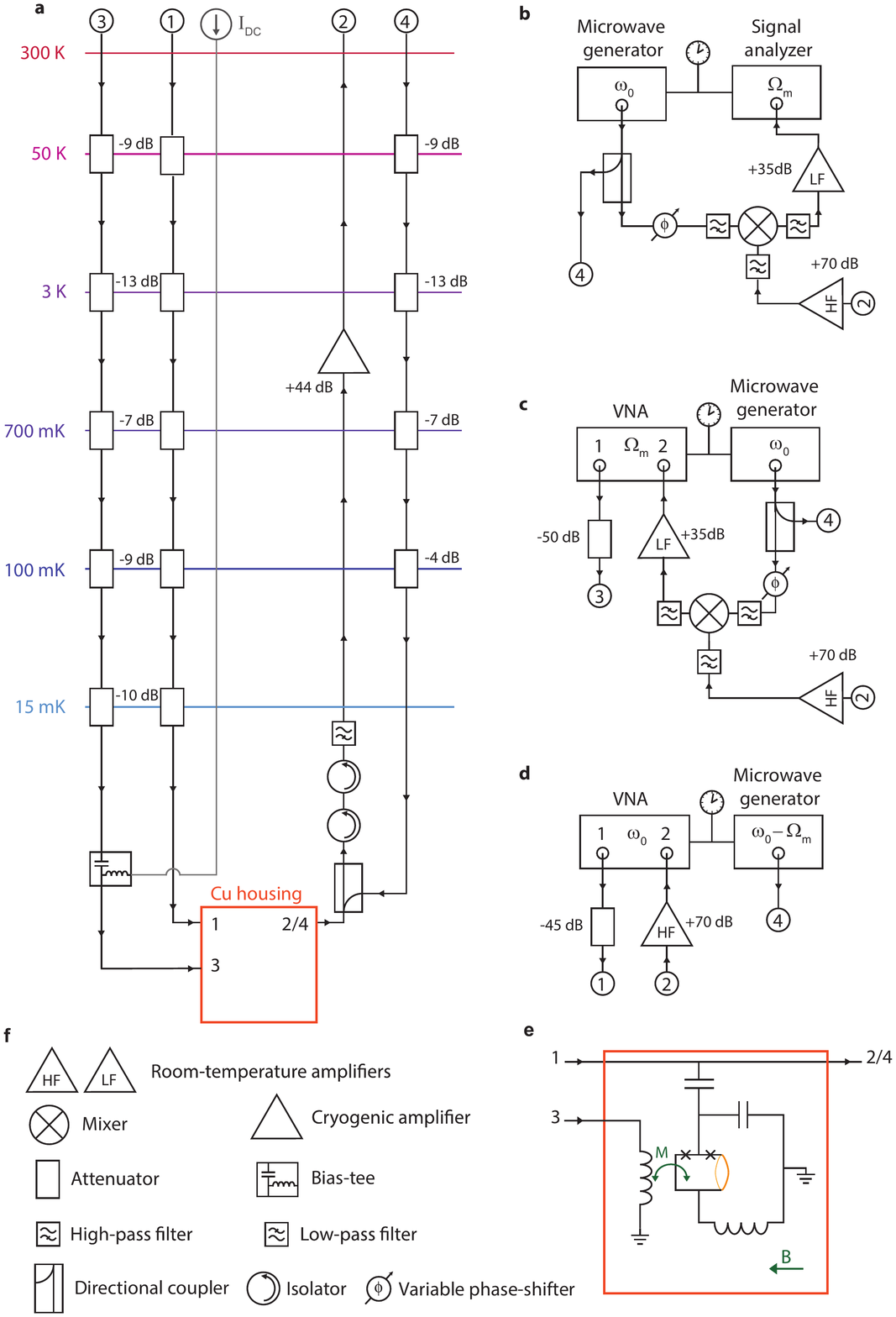}}
		\caption{\textsf{\textbf{Schematic of the measurement setup.} Detailed information is provided in text.}}
		\label{fig:Setup}
	\end{figure}
	All the experiments reported in this paper were performed in a dilution refrigerator operating at a base temperature $T_b \approx 15\,$mK.
	A schematic of the experimental setup and of the external configurations used in the different experiments can be seen in Fig.~\ref{fig:Setup}.
	The PCB (Printed Circuit Board) onto which the fabricated sample was mounted, was placed in a radiation tight copper housing and connected to three high frequency coaxial lines.
	For a rudimentary shielding of magnetic out-of-plane noise without impacting significantly the in-plane magnetic field, a thin superconducting Aluminum cover was placed in parallel $\sim1\,$mm above the chip (not shown in the schematic).
	Two of the coaxial lines were used as standard input and output microwave lines, used to measure the SQUID cavities in a side-coupled transmission configuration.
	Furthermore, in order to generate an out-of-plane magnetic field component, required for the tuning of the SQUID, and for the Lorentz-force actuation of the mechanical resonator, DC currents and low-frequency (LF) signals were sent via a third input line.
	To combine the DC and the LF signals, the center conductor of the coaxial cable was connected to a DC wire by means of a bias-tee.
	All coaxial input lines were heavily attenuated in order to balance the thermal radiation from the line to the base temperature of the refrigerator. 
	Outside of the refrigerator, we used different configurations of microwave signal sources and high-frequency electronics for the three experiments.
	A representation of the setups can be seen in Figs.~\ref{fig:Setup}\textbf{b}, \textbf{c} and \textbf{d}, where the setup for the thermomechanical noise detection is shown in \textbf{b}, the setup for the up-conversion of mechanical motion in \textbf{c}, and the setup for optomechanically induced transparency is shown in \textbf{d}.
	A detailed schematic of the connections inside the cooper housing box is shown in \textbf{e}, and the symbol legend is given in \textbf{e}.
	For all experiments, the microwave sources and vector network analyzers (VNA) as well as the spectrum analyzer used a single reference clock of one of the devices.

	\subsection{Estimation of the attenuation chain}

	To estimate the microwave power on the on-chip microwave feedline, we follow two distinct approaches.
	First, we add all specified loss elements like attenuators or directional couplers.
	Then, we estimate the total additional losses induced by non-specified components like cables and connectors based on a transmission measurement and attribute those additional losses to input and to output cabling, giving significantly more weight to the input lines due to the longer input cables with more potentially lossy connectors.
	For the probe signal line 1, we measure an average transmission of $-5\,$dB, when having $45\,$dB room-temperature attenuation, $48\,$dB cryogenic attenuation, $44\,$dB cryogenic gain at the HEMT amplifier and $70\,$dB gain by room-temperature amplifiers.
	This leaves about $26\,$dB of unaccounted losses, of which we attribute about $17\,$dB to the input and $9\,$dB to the output line, respectively.
	In total, this corresponds to an input attenuation of $-110\,$dB.
	Assuming a similar procedure for the pump input line (line number 2), we get a total attenuation of $\sim -68\,$dB there.
	As second approach, we consider the thermal noise of the HEMT amplifier as calibration standard.
	The HEMT noise power can be determined by
	\begin{equation}
	P_\mathrm{HEMT} = 10\log{\left(\frac{k_\mathrm{B}T_\mathrm{HEMT}}{1\,\mathrm{mW}}\right)} + 10\log{\left(\frac{\Delta f}{\mathrm{Hz}}\right)}
	\end{equation}
	where $k_\mathrm{B}$ is the Boltzmann constant, $T_\mathrm{HEMT}$ is the HEMT noise temperature and according to the data sheet is $T_\mathrm{HEMT} \approx 2\,$K.
	The measurement IF bandwidth of our calibration measurement is $\Delta f = 1\,$kHz.
	In total, we get with these numbers $P_\mathrm{HEMT} = -165.6\,$dBm or the corresponding noise RMS voltage $\Delta V = 1.66\,$nV.
	From the signal-to-noise ratio of SNR$= 34.2\,$dB for a VNA output power of $-20\,$dBm, we then get the signal power arriving at the HEMT input as $-131.4\,$dBm.
	Assuming an attenuation between the sample and the HEMT of $2\,$dB leaves us with a total input attenuation between VNA output and sample of $-109.4\,$dB.
	When performing this procedure with the pump line, we get about $-66\,$dB of attenuation.
	For the calibration of the photon numbers in this paper we therefore work with the attenuations $G_\mathrm{signal} = -110\,$dB and $G_\mathrm{pump} = -67\,$dBm in good agreement with both methods and estimate the accuracy of the achieved calibration on the order of $3\,$dB.
	Note that in addition to the uncertainty mentioned here, the power arriving on the chip is also frequency dependent, as we usually observe background transmission oscillations of about $2\,$ dB peak-to-peak amplitude due to cable resonances.

	\section{Cavity characterization}
	
	\subsection{Cavity modelling}
	
	\subsubsection{Interdigitated capacitors}

	The two interdigitated capacitors $C$ of our device consist of $N = 120$ fingers each, with finger and gap widths of $1\,\mu$m and a finger length $l_f = 100\,\mu$m.
	With the relative permittivity $\epsilon_r = 11.8$ of the Silicon substrate and using the equations given in Ref.~\cite{Igreja04} we calculate the capacitance of one of the main interdigitated capacitors to be $C = 680\,$fF and the interdigitated part of the coupling capacitor as $C_c' = 27\,$fF.
	For the total coupling capacitance, we also have to take into account the capacitance between the center conductor of the feedline and the fingers of both cavity capacitors $C$.
	We do this by calculating the feedline capacitance per unit length $C' = 144\,$pF/m and with a total length of $204\,\mu$m we get $29\,$fF.
	The capacitance between the center conductor and the cavity center electrode, however, is only approximately a factor of $0.25$ of that, such that $C_c = 34\,$fF.
	The resonance frequency of the circuit is $\omega_0 = 2\pi\cdot 5.221\,$GHz and related to the circuit parameters by
	\begin{equation}
	\omega_0 = \frac{1}{\sqrt{L_\mathrm{tot}({2C+C_c})}}
	\end{equation}
	where the total inductance $L_\mathrm{tot} = (L + L_J)/2$.
	The linear inductance $L$ is a combination of the SQUID loop inductance $L_l$ and other linear inductance contributions in the circuit.
	All those have a geometric and a kinetic contribution and from the SONNET simulations discussed below, we estimate the kinetic contribution of the linear inductance to be $L_k = 0.73 L$.
	The total inductance is approximately $L_\mathrm{tot} = 666\,$pH.
	This value is in good agreement with numbers we got using numerical inductance calculation of the whole device and assuming a London penetration depth $\lambda_\mathrm{L} = 160\,$nm, which corresponds to $L_k \approx 2.75 L_g$.

	\subsubsection{SONNET simulations and the kinetic inductance}

	We simulated the cavity with the software package SONNET to determine the kinetic inductance per square $L_\Box$.
	For a vanishing surface impedance we find a resonance frequency $\omega_{00} = 2\pi\cdot 10.05\,$GHz and achieve high agreement with the experimental value of $\omega_0 = 2\pi\cdot 5.221\,$GHz when $L_\Box =  2.3 \,$pH/sq.

	\subsubsection{Analytical cavity model}

\begin{figure}
		\centerline {\includegraphics[trim={0cm 11cm 0cm 0cm},clip=True,width=0.98\textwidth]{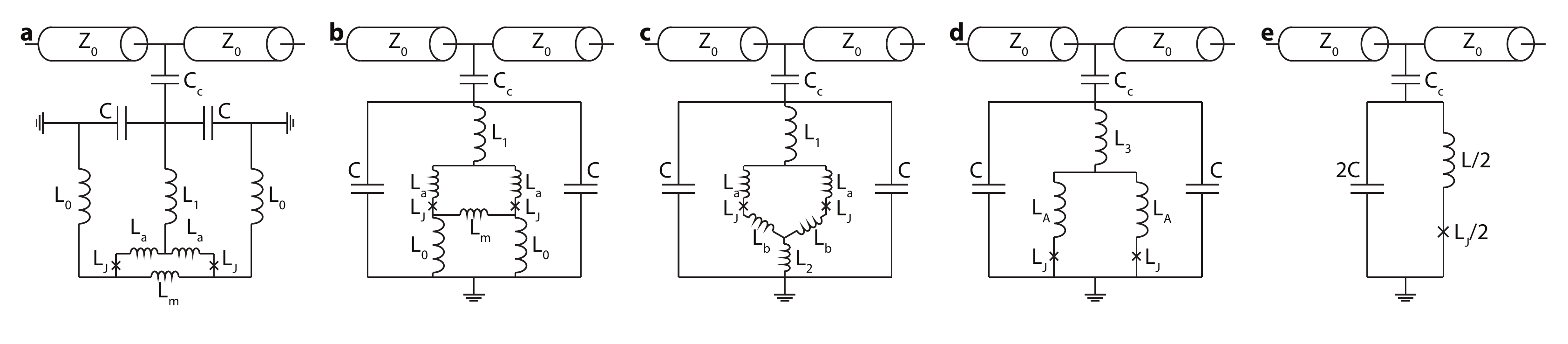}}
		\caption{\textsf{\textbf{Schematic of the device circuit and its simplification.} \textbf{a} The device equivalent circuit with individual circuit elements for each geometric element. \textbf{b} Re-arrangement of the circuit elements with a single ground connection. \textbf{c} Transformation of the inductors $L_0, L_m$ to $L_2, L_b$ using the $\Delta-$Y-approach for impedance-bridges. \textbf{d} Combining series inductors into single inductors $L_3$ and $L_\mathrm{A}$. \textbf{e} Combining parallel elements to get the reduced circuit equivalent.}}
		\label{fig:model}
	\end{figure}

	The cavity used in this experiment is a lumped element SQUID cavity capacitively coupled to a transmission line through a coupling capacitor $C_c$.
	Figure~\ref{fig:model} shows a circuit equivalent of the cavity including the coupling capacitor and the feedline with characteristic impedance $Z_0$.
	In \textbf{a}, a circuit equivalent is shown, which resembles the geometric cavity elements.
	To get a simplified circuit, we first transform the inductances $L_m, L_0$ forming an inductance-bridge via the $\Delta-$Y-approach to the new equivalent inductors
	\begin{eqnarray}
	L_b & = & \frac{L_0 L_m}{2L_0 + L_m}\\
	L_2 & = & \frac{L_0^2}{2L_0 + L_m}
	\end{eqnarray}
	and then combine series and parallel elements to arrive with the simple circuit equivalent shown in Fig.~\ref{fig:model}~\textbf{e}.
	The additional relations between the inductors given in \textbf{a} and \textbf{e} are given by
	\begin{eqnarray}
	L & = & L_A +2L_3\\
	L_A & = & L_a + L_b\\
	L_3 & = & L_1 + L_2.
	\end{eqnarray}

	As values for our device we estimate $L_0 = 1\,$nH, $L_1 = 140\,$pH, $L_m = 60\,$pH and $L_a = 45\,$pH.
	We estimate the critical currents of our Josephson junctions $I_c = 25\,\mu$A, which corresponds to a Josephson inductance of $L_J = 13\,$pH.
	Thus, the total inductance of the circuit is given by $L_\mathrm{tot} = (L+L_J)/2$ and the total capacitance by $C_\mathrm{tot} = 2C + C_c$.

	\subsubsection{Characteristic feedline impedance and external linewidth}

	The external linewidth $\kappa_e$ of the circuit shown in Fig.~\ref{fig:model}\textbf{e} is given by
	\begin{equation}
	\kappa_e = \frac{\omega_0^2C_c^2Z_0}{2C_\mathrm{tot}}
	\end{equation}
	which for our device and a feedline impedance of $Z_0 = 50\,\Omega$ predicts $\kappa_e = 2\pi\cdot3.5\,$MHz.
	This is in slight disagreement with the experimentally determined linewidth of about $9\,$MHz around the flux sweetspot, which can be explained by a combination of two effects.
	First, the on-chip feedline was designed to have a geometric characteristic impedance $Z_{0g} = 50\,\Omega$, which is considerably increased due to the high kinetic inductance of the thin Aluminum film.
	And secondly, we have strong cable resonances in the setup on the order of $3\,$dB peak-to-peak amplitude.
	Both effects considerably modify the effective impedance attached to the circuit.
	When the cavity resonance frequency is tuned and moves through the cable resonances, we also find that the (external) linewidth considerably reduces to about $\kappa = 2\pi\cdot4\,$MHz.
	We found a similarly strong effect with the circuit simulation package QUCS (Quite Universal Circuit Simulator) when introducing cable resonances of similar magnitude as those present in our setup.

	\subsubsection{Intracavity photon number}
	
	The photon number in the cavity is estimated using 
	\begin{eqnarray}
	n_c = \frac{2P_\mathrm{in}}{\hbar\omega_d}\frac{\kappa_e}{\kappa^2 + 4\Delta^2},
	\end{eqnarray}
	where $P_\mathrm{in}$ is the input power (in Watt) on the feedline, $\omega_d$ is the frequency of the drive tone and $\Delta = \omega_d - \omega_0$ the detuning from the cavity resonance.
	Note, that we use $\kappa_e = \kappa$ for this estimation as the device is highly overcoupled.

	\subsection{Response function and fitting routine}
	
	\subsubsection{The ideal cavity response function}

	The $S_{21}^\mathrm{ideal}$ response function of a parallel LC circuit capacitively side-coupled to a transmission line is given by
	\begin{eqnarray}
	S_{21}^\mathrm{ideal} = 1- \frac{\kappa_e}{\kappa_i+\kappa_e+2i\Delta}
	\end{eqnarray}
	with internal and external decay rates
	\begin{eqnarray}
	\kappa_i = \frac{1}{RC_\mathrm{tot}}, \hspace{.3in} \kappa_e = \frac{\omega^2_0C_c^2Z_0}{2C_\mathrm{tot}}
	\end{eqnarray}
	and detuning from the resonance frequency
	\begin{eqnarray}
	\Delta = \omega - \omega_0,\hspace{.3in} \omega_0 = \frac{1}{\sqrt{L_\mathrm{tot}C_\mathrm{tot}}} 
	\end{eqnarray}

	\subsubsection{The real cavity response function}

	The presence of attenuation, cable resonances and parasitic transmission channels is usually captured by additional terms added and multiplied to the ideal cavity response function
	\begin{equation}
	S_{21} = A\left(S_{21}^\mathrm{ideal} + Be^{i\beta}\right)e^{i\alpha}
	\end{equation}
	where $A, B, \alpha, \beta$ are possibly frequency-dependent factors changing the overall transmission function. 
	This can also be written as
	\begin{equation}
	S_{21} = P\left(1 - \frac{Ke^{i\theta}}{\kappa + 2i\Delta}\right)e^{i\phi}
	\label{eqn:S21real}
	\end{equation}
	where $K$ and $\theta$ are functions of $\kappa_e, B$ and $\beta$ and $P$ and $\phi$ are functions of $A, B, \alpha$ and $\beta$.
	Equation (\ref{eqn:S21real}) is used throughout this work for fitting the cavity response function and to extract the total linewidth and resonance frequency.
	Note that a reliable extraction of external and internal linewidths is not possible anymore in the presence of cable resonances and parasitic transmission channels.

	\subsubsection{Full cavity fitting routine}
	\label{sec:BGFitting}

	\begin{figure}
		\centerline {\includegraphics[trim={0cm 6.5cm 0cm 0cm},clip=True,width=0.8\textwidth]{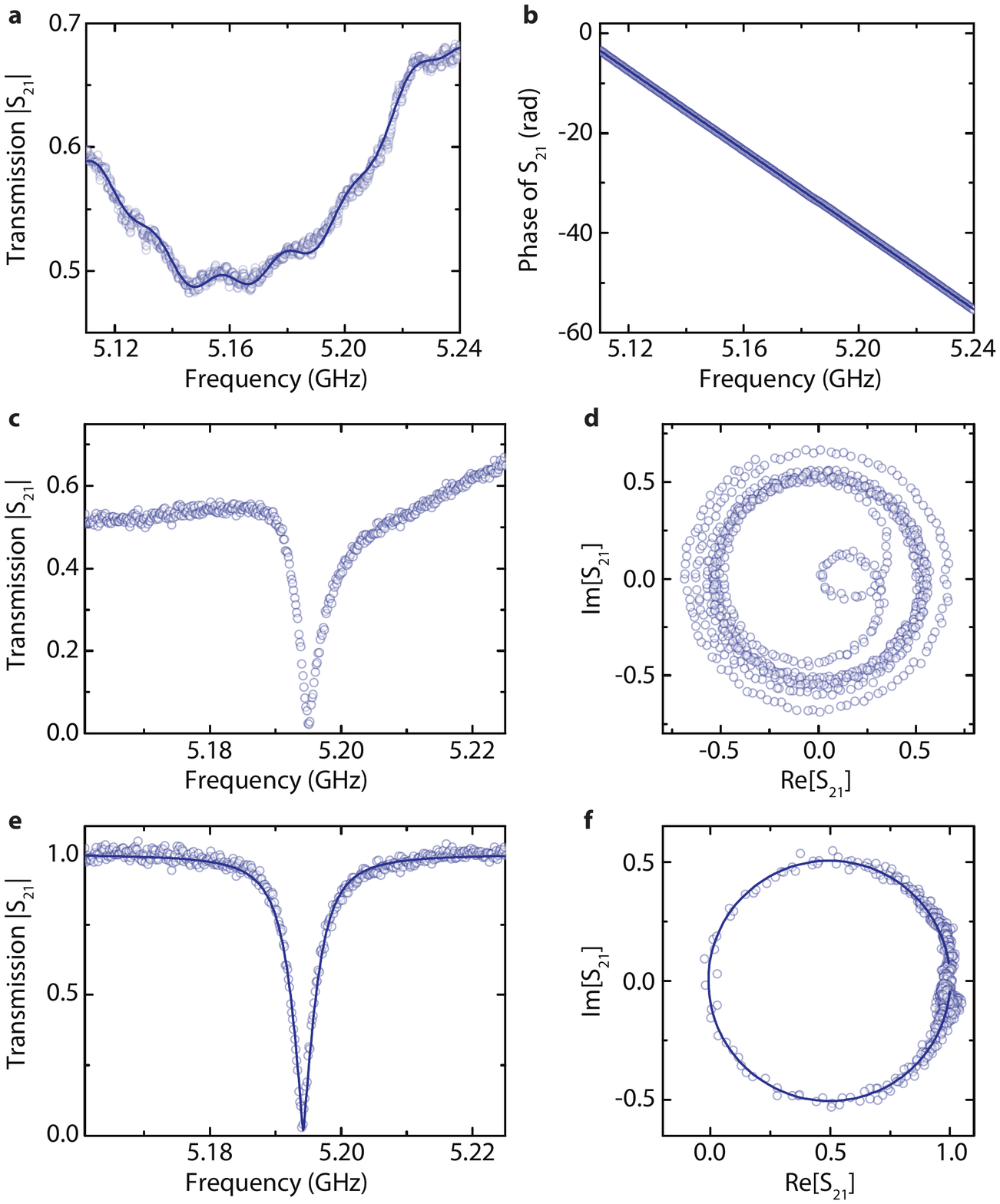}}
		\caption{\textsf{\textbf{The background transmission and how we correct for it.} \textbf{a} Background transmission signal amplitude in the relevant frequency range obtained by tuning the cavity to the maximum and minimum frequency and stitching together the unperturbed parts of the background signal. The corresponding phase is shown in \textbf{b}. Circles show measurement data, lines are fits as described in the text. \textbf{c} and \textbf{d} show the response signal of the cavity in raw data. The amplitude is shown in \textbf{c} and the complex response in \textbf{d}. By means of complex division, we divide off the background fit curves obtained from \textbf{a} and \textbf{b}. The resulting curve is fitted by Eq.~(\ref{eqn:Response}). After this fit, we rotate and rescale the cavity resonance and obtain the signal shown in \textbf{e} and \textbf{f} as circles. The lines show the accordingly rescaled and rotated fits.}}
		\label{fig:Fitting}
	\end{figure}

	During the experiments, the transmitted signals suffer from interferences and losses due to the presence of microwave elements such as attenuators, circulators and amplifiers in the lines, cf. Fig.~\ref{fig:Setup} as well as additional losses from microwave cables.
	For fitting and calibrating the transmitted fields, we follow a step-by-step fitting routine, which is described as follows.
	First, we consider the presence of a frequency dependent background signal expressed as
	\begin{eqnarray}
	S_\mathrm{back} = P(\omega)e^{i\phi(\omega)}.
	\end{eqnarray}

	For the experimental extraction of the background curve, the cavity is initially set to two distant flux bias points with frequencies $\omega_1 = 2\pi\cdot 5.15\,$GHz and $\omega_2 = 5.22\,$GHz and afterwards the spectrum is reconstructed by combining the individual parts where the cavity is non-resonant.
	The amplitude and phase data obtained by this procedure are shown in Fig.~\ref{fig:Fitting}\textbf{a} and \textbf{b} as circles.
	Then, we fit the whole background with a complex function whose magnitude and phase are written as a function of frequency as
	\begin{equation}
	P(\omega) = a_p\omega^5 + b_p\omega^4 + c_p\omega^3 + d_p\omega^2 + e_p\omega + f_p + a_{1c}\cos(b_{1c}\omega + c_{1c}) + a_{2c}\cos(b_{2c}\omega + c_{2c})
	\end{equation}
	\begin{equation}
	\phi(\omega) = a_\phi\omega + b_\phi,
	\end{equation}
	i.e., we perform a linear fit to the phase and both a polynomial and cosine fit to the magnitude of the stitched background data.
	The corresponding fits are shown as lines in Fig.~\ref{fig:Fitting}\textbf{a} and \textbf{b}.
	A measured transmission spectrum with the cavity resonance included is shown in Fig.~\ref{fig:Fitting}\textbf{c} and \textbf{d}.
	Prior to all cavity fits and fits of optomechanically induced transparency, we remove the reconstructed background signal from the measured signal by complex division.
	Considering the possibility that the measured signal might still be influenced by a small frequency-dependent background modulation, we fit the resulting cavity line with
	\begin{eqnarray}
	S_{21} = (a_{p2} + b_{p2}\omega)\left(1-\frac{Ke^{i\theta}}{\kappa+2i\Delta}\right)e^{i(a_{\phi2}\omega + b_{\phi_2})}
	\label{eqn:Response}
	\end{eqnarray}
	where we consider once more a background using the complex scaling factor
	\begin{eqnarray}
	S_\mathrm{back2} = (a_{p2} + b_{p2}\omega)e^{i(a_{\phi2}\omega + b_{\phi_2})}.
	\end{eqnarray}

	Figures~\ref{fig:Fitting}\textbf{e} and \textbf{f} show a resonance curve of the SQUID cavity after background division and rotation by the obtained value of $\theta$ including the cavity response fit using Eq.~(\ref{eqn:Response}).

	\subsection{The Josephson junctions and the SQUID}
	
	\subsubsection{The junctions}

	The constriction type Josephson junctions in our SQUID are designed to be $50\,$nm wide and $200\,$nm long nanobridges in between two superconducting pads, similar to what has been investigated previously by other authors \cite{LevensonFalk11}.
	The pads and the junctions have a constant film thickness of about $20\,$nm and thus we have what is referred to as 2D SQUID geometry in literature \cite{Vijay10}.
	We estimate the critical current to be approximately $I_{c0} \approx 25\,\mu$A.
	Although our junctions might show deviations from an ideal sinusoidal current-phase relation \cite{Vijay10}, we can estimate the zero-bias junction inductance from the critical current to be
	\begin{equation}
	L_{J} = \frac{\Phi_0}{2\pi I_c} \approx 13\,\mathrm{pH}.
	\end{equation}

	\subsubsection{The SQUID loop inductance}

	Due to the 2D SQUID geometry as well as the large kinetic inductance of our films, we have to consider a significant loop inductance when treating the SQUID. 
	From our estimations above, the loop inductance is approximately given by $L_l = 2L_a + L_m \approx 150\,$pH, which gives for the so-called screening parameter
	\begin{equation}
	\beta_L = \frac{2I_{c0}L_l}{\Phi_0} \approx 3.7.
	\end{equation}

	A screening parameter $\beta_L > 2/\pi$ is related to a hysteretic flux state of the SQUID and allows the SQUID to screen more than half a flux quantum before the critical current of the junctions is exceeded by the screening current \cite{Kennedy19a}.

	\subsubsection{Bias flux dependence of the resonance frequency}

	Both, non-sinusoidal current-phase relationship in the form of skewed sine functions as well as large screening parameters lead to widening of the magnetic flux arch and to hysteretic switching of the SQUID flux state.
	Both descriptions have been used to model the hysteretic resonance frequency flux archs of superconducting resonators including SQUIDS with constriction type Josephson junctions \cite{LevensonFalk11, Kennedy19a}.
	We phenomenologically include both effects in the description of the single-arch flux-dependence of our SQUID cavities by including a factor $\gamma_L$ into the effective single junction inductance
	\begin{eqnarray}
	L_J(\Phi) = \frac{L_{J0}}{\cos\left( \pi\gamma_L \frac{\Phi}{\Phi_0} \right)}.
	\end{eqnarray}

\begin{figure}
		\centerline {\includegraphics[trim={0cm 21.5cm 0cm 0cm},clip=True,width=0.9\textwidth]{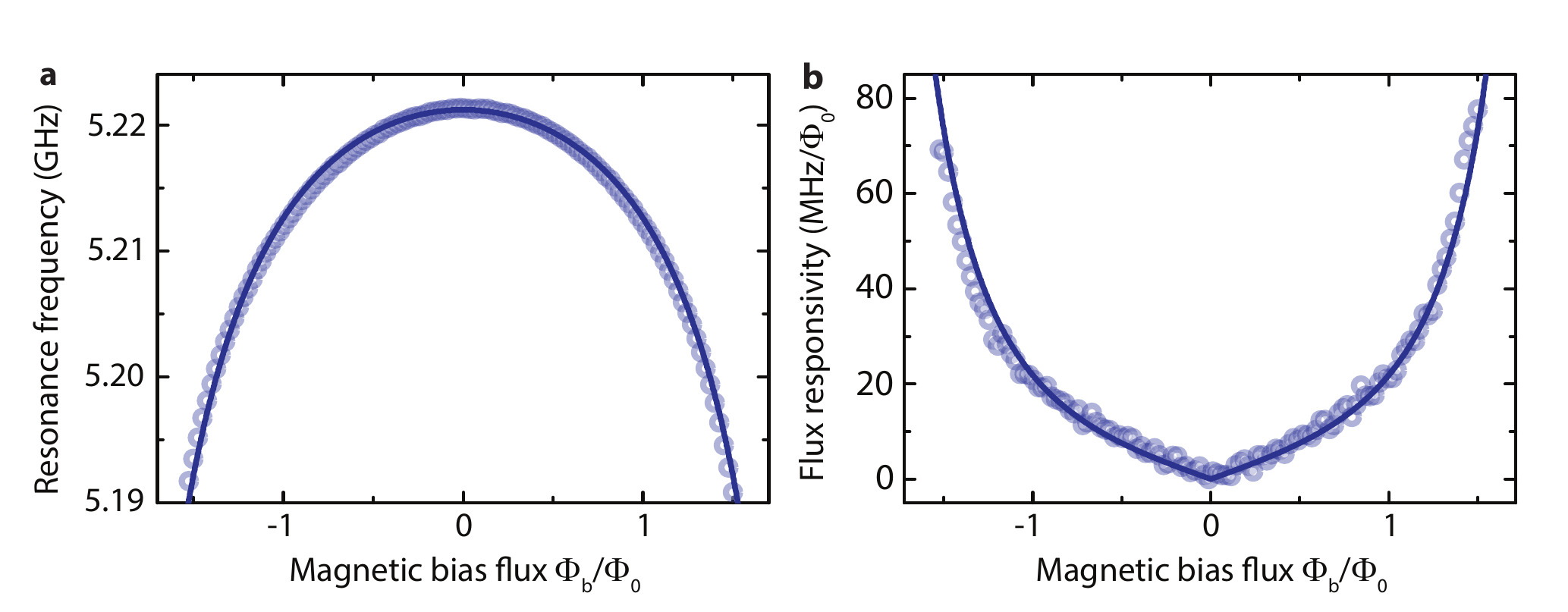}}
		\caption{\textsf{\textbf{Cavity frequency tuning and flux responsivity with magnetic bias flux.} \textbf{a} Cavity resonance frequency vs magnetic bias flux for $B_{||} = 3\,$mT. Circles are data points extracted from fits and the line is a fit using Eq.~(\ref{eqn:Arch}) with fixed $\Lambda = 0.99$ and $\gamma_\mathrm{L}$ being the only free parameter. In \textbf{b} the flux responsivity $\left|\partial\omega_0/\partial\Phi\right|$ is plotted. Both, the experimental and the theoretical curves are obtained by calculating the derivative of the data in \textbf{a}.}}
		\label{fig:FluxArch}
	\end{figure}

	The factor $\mathrm{\gamma_L}$ takes a widening of the flux arch and a tuning of the resonance frequency far beyond $\pm\Phi_0/2$ into account, cf. Fig.~\ref{fig:FluxArch}, where an ideal SQUID with a sinusoidal current-phase relationship and negligible loop inductance would have $\gamma_L = 1$.
	The resonance frequency of the SQUID cavity can therefore be expressed as
	\begin{eqnarray}
	\omega_0(\Phi) = \frac{1}{\sqrt{C_\mathrm{tot}(L + L_J(\Phi))/2}}.
	\end{eqnarray}

	Defining the sweet spot resonance frequency by
	\begin{equation}
	\omega_0^s = \frac{1}{\sqrt{C_\mathrm{tot}(L+L_{J0})/2}}
	\end{equation}
	we can write the flux-dependent frequency as
	\begin{equation}
	\omega_0(\Phi) = \frac{\omega_0^s}{\sqrt{\Lambda +\frac{1-\Lambda}{\cos{\left(\pi\gamma_L\frac{\Phi}{\Phi_0}\right)}}}}.
	\label{eqn:Arch}
	\end{equation}
	with $\Lambda = L/(L+L_{J0})$.
	For our device parameters, we get $\Lambda \approx 0.99$.
	Figure~\ref{fig:FluxArch}\textbf{a} shows the resonance frequency of the SQUID cavity when biased with the on-chip bias line and the resulting flux arch was fitted with Eq.~(\ref{eqn:Arch}).
	The only free parameter for the fit was $\gamma_L = 0.23$, indicating a large screening parameter and/or a non-sinusoidal current-phase relation.
	We note here, however, that the theoretical $\beta_L = 3.7$ derived above is too small to explain the widening of the arch as we observe it.
	One possible explanation is a non-sinusoidal current-phase relation.
	A second possibility, which is at the same time in agreement with the deviation between theory and experiment of the mechanical resonance frequency shift with in-plane field, is that we underestimate the loop inductance significantly.
	A discussion of this possibility with a possible explanation is given in Sec.~\ref{sec:dOmega}.
	In Fig.~\ref{fig:FluxArch}\textbf{b}, we plot the derivative of both, the data points and the fit curve, to obtain the flux responsivity $\partial\omega_0/\partial\Phi$, which is directly proportional to the optomechanical single-photon coupling rate $g_0$.
	Both parameters, $\Lambda$ and $\gamma_L$ seem to depend slightly on the magnetic in-plane field, which is taken into account in our analysis.
	The values given here are extracted for $B_{||} = 1\,$mT.
	The origin of this dependence, however, is not fully clear.
	It might be due to a change of the bias current flow for large in-plane fields or to a change of kinetic loop inductance, while the rest of the kinetic cavity inductance stays nearly unchanged.

	\subsubsection{Calibration of the flux axis}

	We use a combination of the measured $\Phi_0$-periodicity of jumps in the hysteretic resonance frequency, when the bias flux is changed and simulations of the biasing to calibrate the flux axis for the bias flux dependence.
	\begin{figure}
		\centerline {\includegraphics[trim={0.5cm 12.5cm 0.5cm 8cm},clip=True,width=0.8\textwidth]{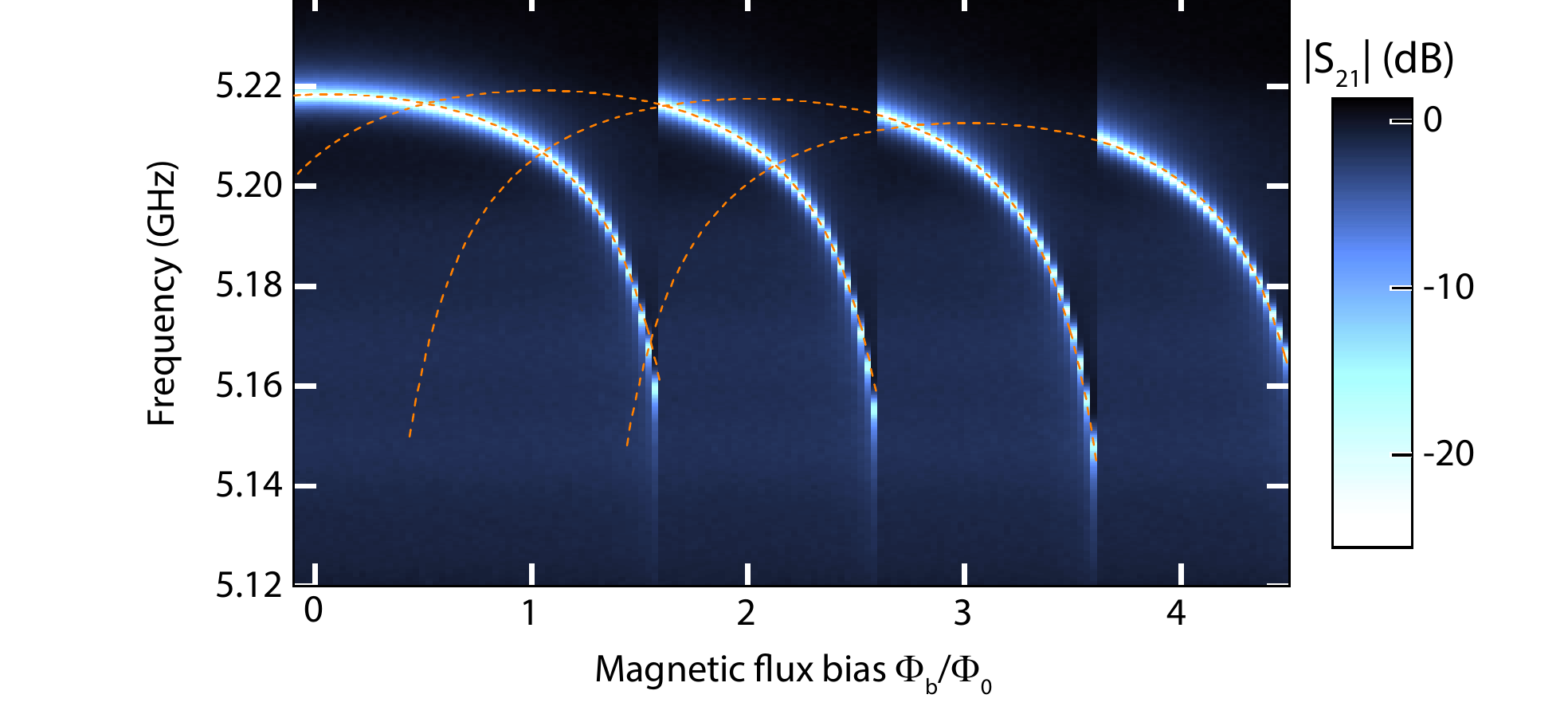}}
		\caption{\textsf{\textbf{Cavity frequency tuning with magnetic bias flux beyond a single flux arch.} When we sweep the bias flux to larger values than about $1.6\Phi_0$, we find periodic jumps in the resonance frequency and partial archs. This is an indication for a non-negligble screening parameter and/or a non-sinusoidal current-phase relation. The periodicity can be used to calibrate the flux axis to the flux quantum $\Phi_0$. The dashed lines correspond to copies the flux arch dependence used in Fig.~\ref{fig:FluxArch}\textbf{a} each shifted in flux and sweetspot frequency only to match the observed resonance frequencies.}}
		\label{fig:Jumps}
	\end{figure}
	Figure~\ref{fig:Jumps} shows an example for the hysteretic jumps of the cavity frequency with flux, indicating a significant loop inductance and/or a non-sinusoidal current-phase relation \cite{LevensonFalk11, Kennedy19a}.
	Under the assumption, that $I_{c0}$ and $L_l$ are not significantly depending on the magnetic in-plane field, we herewith calibrate the flux axis for all in-plane fields.
	This assumption is justified by a nearly constant sweetspot resonance frequency of the cavity and by a nearly constant periodicity between the jumps with respect to the biasing current.
	Note, that in contrast to the description given in Ref.\cite{Kennedy19a}, the periodicity of the jumps corresponds to $1\Phi_0$ instead of $2\Phi_0$.

	\subsubsection{Cavity anharmonicity}

	Assuming a sinusoidal current-phase relation, we calculate the shift per photon to first order by
	\begin{equation}
	\chi = -\frac{e^2}{2\hbar C_\mathrm{tot}}\left(1-\Lambda\right)^3 \approx 2\pi\cdot 14\textbf{}\,\mathrm{Hz}.
	\end{equation}

	Therefore, the cavity can be considered in good approximation as linear, as long as the photon number does not exceed a few 1000.

	\section{Mechanical characterization}
	
	\subsection{Theory of Lorentz-force actuation}

	The equation of motion of the mechanical resonator is given by
	\begin{equation}
	\ddot{x} + \frac{\Omega_0}{Q_m}\dot{x} + \Omega_0^2 x = \frac{F(t)}{m}
	\end{equation}
	where $m$ is the effective mass, $Q_m$ is the mechanical quality factor and $\Omega_0$ is the resonance frequency.
	External forces onto the mechanical oscillator are contained in $F(t)$.
	The current through the mechanical beam in presence of flux biasing and a magnetic in-plane field is given by the flux quantization and conservation in the SQUID loop.
	In the absence of a bias current and for identical junctions, the general relation between the phase difference across one junction $\delta$ and the total flux through the loop $\Phi$ is given by
	\begin{equation}
	\frac{\delta}{\pi} = \frac{\Phi}{\Phi_0}.
	\end{equation}

	The circulating current at the same time is related to the phase difference by
	\begin{equation}
	J = I_{c0}\sin{\delta}
	\end{equation}

	The total flux through the loop $\Phi$ is a sum of the bias flux $\Phi_b$, the flux generated by a loop current via the loop inductance $\Phi_J = L_lJ$, and a contribution from the in-plane field when the mechanical oscillator is displaced from its equilibrium position $\Phi_x = \gamma B_{||}lx$, thus
	\begin{eqnarray}
	\Phi & = & \Phi_b + \Phi_J + \Phi_x \\
	& = & \Phi_b + L_lJ + \gamma B_{||}lx.
	\end{eqnarray}

	For a constant flux bias $\Phi_{b0}$ there is a circulating current $J_0$ and the mechanical beam is in the equilibrium position $x_0$.
	We assume now that all quantities only slightly differ from their equilibrium values $\Phi_b(t) = \Phi_{b0} - \Delta\Phi_b$, $x(t) = x_0 +\Delta x$ and $J(t) = J_0 + \Delta J$.
	Redefining $x = \Delta x$ and $L_J = L_J(\Phi_{b0})$, we can approximate to first order
	\begin{equation}
	\Delta J = \frac{\Delta\Phi_b}{L_l + 2L_J} - \frac{\gamma B_{||}l x}{L_l + 2L_J}.
	\end{equation}

	The dynamical part of the Lorentz-force is given by $F_\mathrm{L}(t) = \gamma B_{||}l\Delta J$ and thus the equation of motion becomes
	\begin{equation}
	\ddot{x} + \frac{\Omega_0}{Q_m}\dot{x} + \left(\Omega_0^2 + \frac{\gamma^2 B_{||}^2 l^2}{m(L_l+2L_J)}\right) x = \frac{\gamma B_{||}l}{m(L_l+2L_J)}\Delta\Phi_b(t).
	\end{equation}

	Thus, a time-varying magnetic flux is translated into a time-varying Lorentz-force and can be used to directly drive the mechanical motion.
	In addition, a position-dependent force emerges from the mechanical oscillator placed in a SQUID loop, which shifts the mechanical resonance frequency.

	\subsection{In-plane magnetic field dependence}
	\label{sec:dOmega}

\begin{figure}
		\centerline {\includegraphics[trim={0cm 0cm 0cm 1cm},clip=True,width=0.45\textwidth]{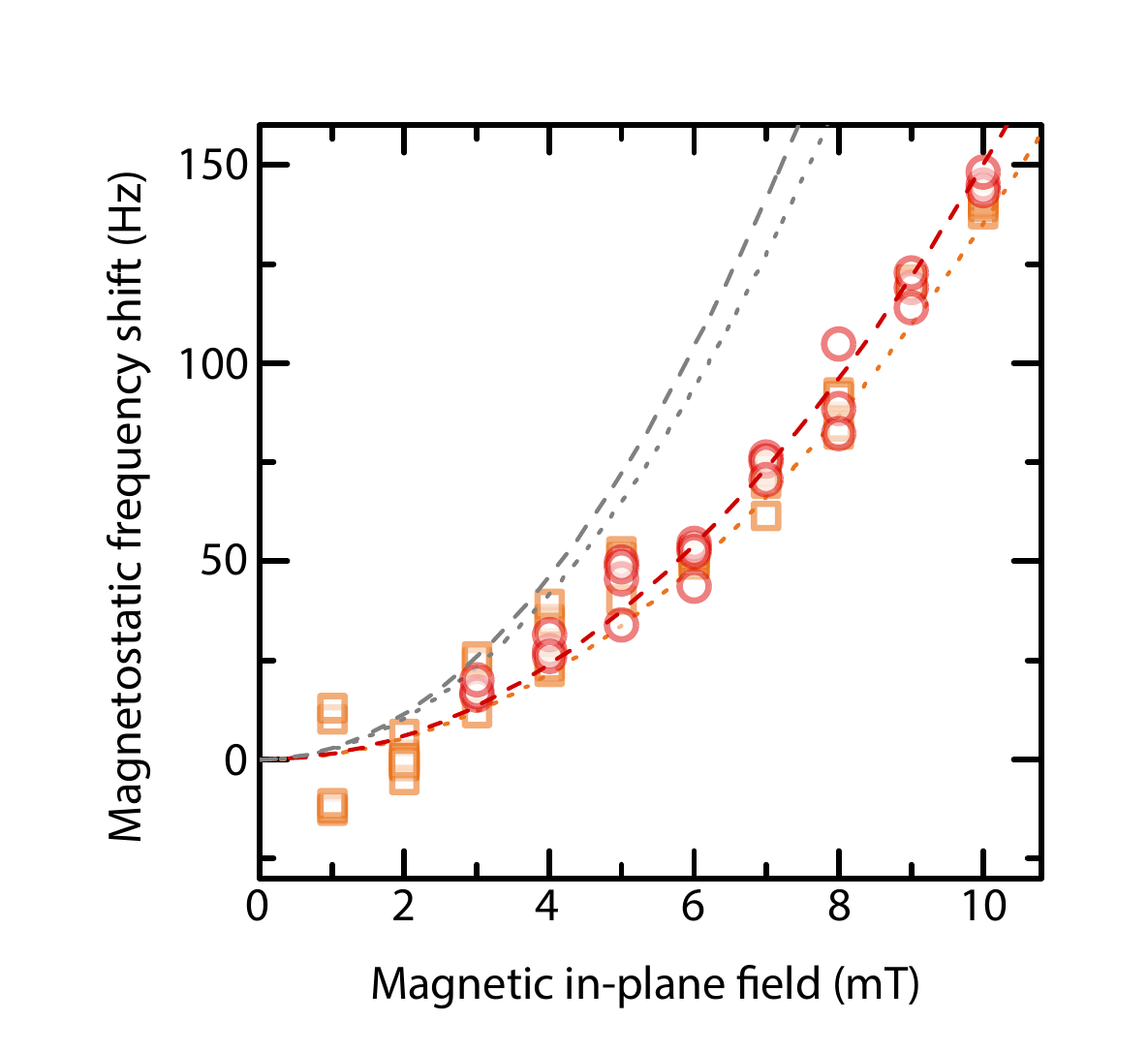}}
		\caption{\textsf{\textbf{Magnetostatic spring stiffening by Lorentz-force backaction.} The measured frequency shift is plotted as points for two different values of bias flux. Circles corrrespond to $\Phi_b/\Phi_0 = 0.75$ and squares to $\Phi_b/\Phi_0 = 1.45$. The gray dashed and dotted lines are the theoretical calculations without free parameters and overestimate the measured effect by a factor of $\sim 2$. The red dashed and orange dotted lines correspond to the theoretical lines with a scaling factor of $\sim 0.52$ and agree well with the observed frequency shift.}}
		\label{fig:Shift}
	\end{figure}

	The position dependent part of the Lorentz-force is equivalent to a mechanical spring stiffening, in analogy to the electrostatic softening in electromechanical capacitors.
	The shifted resonance frequency is given by
	\begin{equation}
	\Omega_m^2 = \Omega_0^2 + \frac{\gamma^2 B_{||}^2 l^2}{m(L_l+2L_J)}
	\end{equation}
	what can be approximated as
	\begin{equation}
	\Omega_m \approx \Omega_0 + \frac{\gamma^2 B_{||}^2l^2}{2m\Omega_0(L_l + 2L_J)}.
	\end{equation}
	We indeed observe a shift of the mechanical resonance frequency with in-plane field as shown in Fig.~\ref{fig:Shift} for two different flux responsivities, i.e., for two different Josephson inductances.
	The absolute numbers, however, are smaller by about a factor of $\sim 2$ than the result of independent calculations based on the device parameters and the in-plane field.
	Possible reasons for this mismatch is the overestimation of the mode scaling factor $\gamma = 0.86$, which we determined through matching the experimental $g_0$ with the theoretical calculations, an underestimated loop inductance or a field-dependent loop or Josephson inductance.
	In combination with the observation of flux arch widening, we consider the most probable explanation that the loop inductance is significantly higher than expected.
	For the mechanical resonance frequency shift, we find a good agreement between theory and experiment for a loop inductance of $\sim 350\,$pH.
	This would correspond to $\beta_L = 8.6$.
	A possible origin for this deviation is possibly related to the suspension of the mechanical part of the loop and the release process, which ends with oxygen plasma ashing of the resist and might induce an enhanced oxidation of the bottom side of the beam.
	The mechanical beam oxidizes from the top and the bottom, while the rest of the circuit only oxidizes from the top.
	For very thin films as used here, the oxide layer of a few nm thickness might change the thickness of the superconducting layer significantly, which will increase the kinetic inductance of that region.
	As the inductance of our circuit is dominated by kinetic inductance anyways, such a two-sided oxidization might indeed be responsible for a significantly increased inductance of the suspended parts.
	This would explain, why the results related to the loop inductance are deviating from theoretical calculations, while all results where the loop inductance is not relevant, are in excellent agreement.

	\subsection{Upconversion of coherently driven mechanical motion}

\begin{figure}
		\centerline {\includegraphics[trim={2cm 0cm 3cm 0.8cm},clip=True,width=0.95\textwidth]{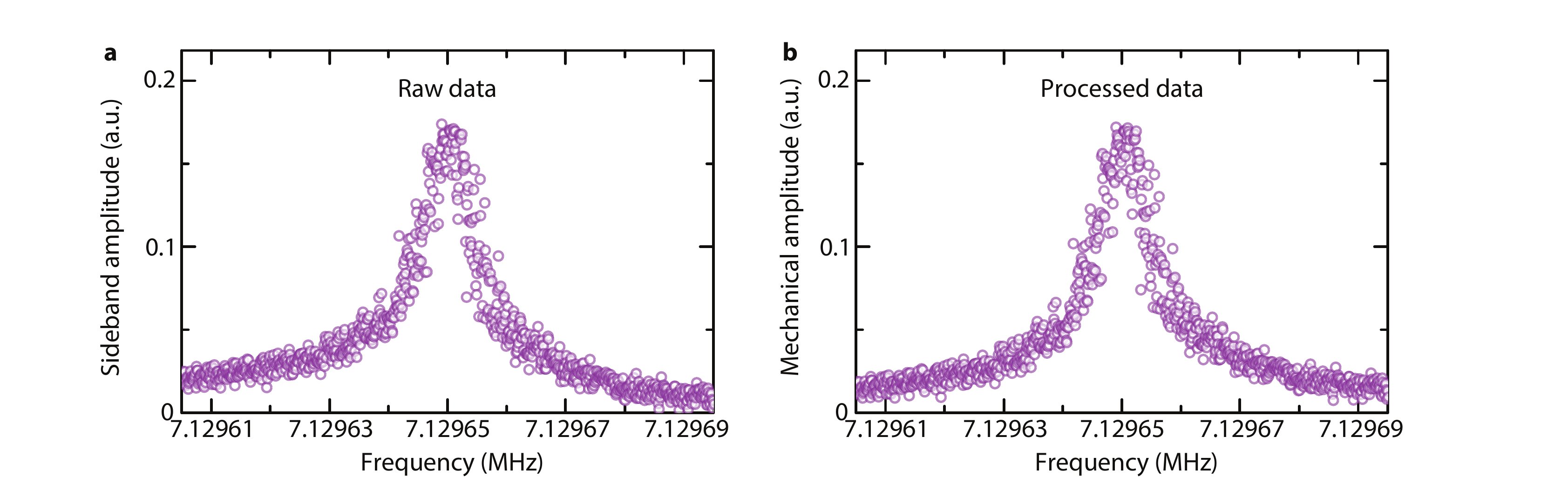}}
		\caption{\textsf{\textbf{Processing the motional sideband generated by mechanical displacement.} \textbf{a} Raw data for the sideband amplitude detected by means of sending a resonant tone into the cavity while exciting the mechanical mode by Lorentz-force. The Lorentz-force drive current frequency is swept through the mechanical resonance. Due to additional contributions to the SQUID cavity sideband such as direct flux modulation of the SQUID by the Lorentz-force current, the sideband does not only contain information about the mechanical displacement. \textbf{b} shows the amplitude data of \textbf{a}, where a constant complex number has been substracted from the complex $S_{21}$ data.}}
		\label{fig:MechUp}
	\end{figure}

	We excite the mechanical resonator by Lorentz-force actuation and measure the cavity sidebands generated by the corresponding cavity field phase modulation when sending a tone resonant with the cavity $\omega_d = \omega_0$.
	The excitation current is generated by the output port of a vector network analyzer and sent through the on-chip bias line, cf. Fig.~\ref{fig:Setup}\textbf{c}.
	At the same time, we drive the cavity with a resonant microwave tone generated by a signal generator.
	The cavity output field, including the motional sidebands, is amplified and sent through a high-pass filter into a mixer, where it is down-converted by being mixed with the original carrier tone.
	The mixer output is low-pass filtered and sent into the input port of the network analyzer.
	As we are driving the cavity on resonance, we must adjust the phase of the carrier signal in order to get constructive interference of the sidebands at $+\Omega$ and $-\Omega$. 
	We adjust the phase-shifter manually until the detected sideband signal is maximized.
	In this setup, however, we do not only detect the additional flux induced into the SQUID by the mechanical motion, but also the phase modulations directly generated by the bias flux modulation itself.
	Other possible parasitic tones come from mixing due to the cavity nonlinearity or in the nonlinear elements of the detection chain.
	The detected sideband amplitude $|S_{21}|$ is thus proportional to
	\begin{equation}
	|S_{21}(\Omega)| \propto \left| \frac{\gamma B_{||} l}{2m\Omega_m} \frac{F_\mathrm{L}(\Omega)}{\Omega_m - \Omega - i\frac{\Gamma_m}{2}} + Se^{i\sigma}\right|
	\end{equation}
	with an additional signal $Se^{i\sigma}$ interfering with the motional sideband.
	Therefore, the measured, upconverted mechanical resonance will have a slight Fano lineshape as shown in Fig.~\ref{fig:MechUp}\textbf{a}.
	We correct for this slight asymmetry by substracting a constant complex number from the detected signal.
	The result is shown in Fig.~\ref{fig:MechUp}\textbf{b} and in Fig.~2 of the main paper.

	\subsection{Interferometric detection of thermal mechanical motion}

	The measurement routine is very similar to the one for the detection of coherently driven motion.
	Instead of using a network analyzer, however, we do not apply any driving current, but just detect the down-converted sideband-voltage quadratures $I$ and $Q$ with a vector signal analyzer.
	From the Fourier-transform of the quadratures, we calculate the corresponding power spectral density.

	\section{Optomechanical device characterization}
	
	\subsection{Optomechanical equations of motion}
	
	The system is modelled with the classical equations of motion for the mechanical displacement $x$ and normalized intracavity field amplitude $\alpha$
	\begin{eqnarray}
	\ddot{x} =  - \Gamma_m\dot{x} -\Omega_m^2x + \frac{1}{m}(F_r + F_e)
	\end{eqnarray}
	\begin{eqnarray}
	\dot{\alpha} = \left[i(\Delta + Gx) - \frac{\kappa}{2}\right]\alpha + \sqrt{\frac{\kappa_e}{2}}S_\mathrm{in} ,
	\end{eqnarray}
	where $\Delta = \omega_d - \omega_0$ is the detuning from the cavity resonance frequency, $\kappa = \kappa_i + \kappa_e$ is the total cavity linewidth and $S_\mathrm{in}$ is the normalized input field.
	The external forces onto the mechanical oscillator are expressed by $F_e$ and the radiation pressure force contribution is taken into account in $F_r$ and expressed as a function of the intracavity field by
	\begin{eqnarray}
	F_r = \hbar G \left| \alpha \right| ^2,
	\end{eqnarray}
	with pull parameter G
	\begin{eqnarray}
	G = -\frac{\partial\omega_0}{\partial x}.
	\end{eqnarray}

	Assuming that the intracavity field is high enough to only consider small deviations from the steady state solutions with $x = \bar{x} + \delta x$ and $\alpha = \bar{\alpha} + \delta \alpha$ and no external driving force $F_e$, the equations of motion can be linearized as
	\begin{eqnarray}
	\delta\ddot{x} =  - \Gamma_m\delta \dot{x} -\Omega_m^2\delta x + \frac{\hbar G \bar{\alpha}} {m}(\delta \alpha + \delta\alpha^*)
	\end{eqnarray}
	\begin{eqnarray}
	\delta\dot{\alpha} = \left[ i\bar{\Delta} - \frac{\kappa}{2}\right]\delta \alpha + iG\bar{\alpha}\delta x + \sqrt{\frac{\kappa_e}{2}}S_{p} 
	\end{eqnarray}

	In the above expressions, the detuning $\bar{\Delta} = \omega_d - \omega_c +G\bar{x}$ takes into account the shift from the equilibrium position $\bar{x}$ due to the radiation pressure force and $\sqrt{\frac{\kappa_e}{2}}S_p$ with $S_p = S_0e^{-i\Omega t}$, $\Omega = \omega - \omega_d$ accounts for field fluctuations.
	As in our experiments $\bar{\Delta} \approx \Delta$, we will just use $\Delta$ instead of $\bar{\Delta}$ throughout this paper.
	The response of the optomechanical cavity is then given by
	\begin{equation}
	S_{21} = 1 - \sqrt{\frac{\kappa_e}{2}}\frac{a_{-}}{S_0}
	\end{equation}
	with
	\begin{equation}
	a_{-} = \chi_c \left[ 1+i2m\Omega_mg^2\chi_c\chi^\mathrm{eff}_m \right] \sqrt{\frac{\kappa_e}{2}}S_0.
	\end{equation}

	Here
	\begin{equation}
	\chi_{c} = \frac{1}{\frac{\kappa}{2}-i(\Delta + \Omega)}
	\end{equation}
	is the cavity susceptibility and 
	\begin{equation}
	\chi^\mathrm{eff}_{m} = \frac{1}{2m\Omega_m}\frac{1}{\Omega_m - \Omega -i\frac{\Gamma_m}{2} + \Sigma(\Omega_m)}
	\end{equation}
	with 
	\begin{eqnarray}
	\Sigma(\Omega_m) & = & -ig^2\left[ \chi_c(\Omega_m)-\chi^{*}_c(-\Omega_m) \right]
	\label{eqn:Sigma}
	\end{eqnarray}
	is the effective mechanical susceptibility in the high-$Q_m$ approximation.

	\subsection{Optical spring and optical damping}

	By re-writing Eq.~(\ref{eqn:Sigma}) as $\Sigma = \delta\Omega_m  - i\Gamma_0/2$  and analyzing the real and imaginary part we can write the change in mechanical frequency $\delta\Omega_m$ (optical spring) and the additional mechanical damping term $\Gamma_o$ (optical damping) as
	\begin{eqnarray}
	\delta\Omega_m & = & g^2 \left[ \frac{\Delta + \Omega_m}{\frac{\kappa^2}{4} + (\Delta + \Omega_m)^2} + \frac{\Delta - \Omega_m}{\frac{\kappa^2}{4} + (\Delta - \Omega_m)^2} \right]\\
	\Gamma_o & = & g^2\kappa \left[ \frac{1}{\frac{\kappa^2}{4} + (\Delta + \Omega_m)^2} - \frac{1}{\frac{\kappa^2}{4} + (\Delta - \Omega_m)^2} \right]
	\end{eqnarray}

	For all our experimental parameters, the optical frequency shift is negligibly small $\delta\Omega_m <1\,$Hz, i.e., $\delta\Omega_m \ll \Gamma_m$, and therefore is not accounted for in any of the measurements or analyses.

	\subsection{Optomechanically induced transparency in the unresolved sideband regime}

	For our device, we have $\kappa \sim \Omega_m$ and thus we cannot use the approximate equations and results for the resolved sideband regime.
	We used two related methods to analyze our experiments on optomechanically induced transparency and to determine the single-photon coupling rate $g_0$ from these measurements.
	For a drive on the red sideband and $\Gamma_m \ll \kappa$, both resonances, the cavity response as well as the response window of the mechanical oscillator inside the cavity describe a circle in the complex response.
	The ratio of the diameters of these circles can be used to determine the optomechanical multi-photon coupling rate $g$ as described below.
	In the second way, we fit both resonances with a complex resonance function as Eq.~(\ref{eqn:Response}) and determine the cooperativity from the ratio of the amplitudes on resonance.

	\subsubsection{Cavity circle diameter $d_c$}

	To demonstrate that the circle diameter ratio is not influenced by the presence of parasitic resonances and transmission channels of the setup, we start with the modified optomechanical response function similar to what we described above for the bare cavity
	
	\begin{equation}
	S_{21} = A\left(1-\frac{\kappa_e}{\kappa + 2i(\Delta+\Omega)}\left[1+i2m\Omega_mg^2\chi_c\chi_m^\mathrm{eff}\right] + Be^{i\beta}\right)e^{i\alpha}
	\end{equation}
	which can be rewritten as
	
	\begin{equation}
	S_{21} = P\left(1-\frac{Ke^{i\theta}}{\kappa + 2i(\Delta+\Omega)}\left[1+i2m\Omega_mg^2\chi_c\chi_m^\mathrm{eff}\right]\right)e^{i\phi}.
	\end{equation}

	From the bare cavity fit, we determined the background $S_\mathrm{back} = Pe^{i\phi}$ and we divide this background off to get
	
	\begin{equation}
	S_{21} = 1-\frac{K e^{i\theta}}{\kappa + 2i(\Delta+\Omega)}\left[1+i2m\Omega_mg^2\chi_c\chi_m^\mathrm{eff}\right]
	\end{equation}

	For $|\Omega_m - \Omega| \gg \Gamma_m$, the mechanical susceptibility essentially vanishes in the weak coupling limit and we get back the bare cavity response function 
	
	\begin{eqnarray}
	S_{21} & = & 1 - \frac{Ke^{i\theta}}{\kappa + 2i(\Delta +\Omega)}.
	\end{eqnarray}
	
	By calculating the cavity response at the points $\Omega = -\Delta-\kappa/2$ and $\Omega = -\Delta+\kappa/2$ we get
	
	\begin{eqnarray}
	S_{21-} & = & 1 - \frac{Ke^{i\theta}}{\kappa + i\kappa},\hspace{.3in} S_{21+} = 1 - \frac{Ke^{i\theta}}{\kappa - i\kappa}
	\end{eqnarray}
	
	The distance between these two points gives us the bare cavity circle diameter
	
	\begin{eqnarray}
	d_c = | S_{21-} - S_{21+} | & = &\frac{K}{\kappa}.
	\end{eqnarray}

	\subsubsection{OMIT circle diameter $d_m$}

	For the estimation of the diameter of the circle related to the mechanical signal as optomechanically induced transparency (OMIT), we first consider that the anchor point of the mechanical circle does not necessarily correspond exactly to the cavity resonance frequency in order to account for cases where there is still a small detuning present in the experiment.
	This offset $\delta_m = \omega_0-\omega_d - \Omega_m$ will modify the diameter of the circle with respect to the resonant case.
	Considering $\Gamma_\mathrm{eff} = \Gamma_m + \Gamma_o \ll \kappa$ we can expect that, for a fixed pump frequency close to the the cavity red sideband $\Delta \approx -\Omega_m-\delta_m$, the cavity has a constant reponse during the OMIT circle, given by
	
	\begin{eqnarray}
	\chi_c = \frac{2}{\kappa-2i\delta_m}
	\end{eqnarray}
	
	By evaluating the total response function at the points $\Omega = \Omega_m - \Gamma_\mathrm{eff}/2$ and $\Omega = \Omega_m + \Gamma_\mathrm{eff}/2$ we calculate the OMIT circle diameter
	
	\begin{eqnarray}
	d_m & = & \left| S_{21-} - S_{21+} \right|= \left|-4iKm\Omega_mg^2\frac{1}{(\kappa-2i\delta)^2}\left[ \chi_m^\mathrm{eff-}-\chi_m^\mathrm{eff+}\right] e^{i\theta} \right|\\
	& = & 4K\frac{g^2}{\Gamma_\mathrm{eff}}\frac{1}{\kappa^2 + 4\delta_m^2}.
	\end{eqnarray}

	\subsubsection{Effective cooperativity $C_\mathrm{eff}$ and the extraction of $g_0$}
	
	We define the effective cooperativity as
	
	\begin{equation}
	C_\mathrm{eff} = \frac{4g^2}{\kappa\Gamma_\mathrm{eff}}
	\end{equation}
	
	and with this the ratio of the cavity and mechanical resonance circle diameters is given by
	
	\begin{equation}
	\frac{d_m}{d_c} = C_\mathrm{eff}\frac{\kappa^2}{\kappa^2 + 4\delta_m^2}
	\end{equation}

	Thus, as a measurement of the cavity and the transparency window of OMIT provide us with the circle diameters, the cavity linewidth $\kappa$ and the detuning $\delta_m$, we can extract the effective cooperativity, which in combination with the width of the transparency window $\Gamma_\mathrm{eff}$ allows for the extraction of the multi-photon coupling rate $g$.
	Using the estimated intracavity photon number $n_c$ finally leads to the single-photon coupling rate
	
	\begin{equation}
	g_0 = \frac{g}{\sqrt{n_c}}.
	\end{equation}

	\subsection{Full experimental and fitting procedure for optomechanically induced transparency}
	
	\subsubsection{Adjusting the cavity parameters and measurement routine}

	\textbf{I.} As first step in all measurements, we fix the in-plane field to a desired value $B_{||}$.
	\textbf{II.} As second step, we sweep the bias flux in small steps and for each value take a transmission spectrum of the cavity with a network analyzer.
	The cavity resonance is fitted within the measurement script using Eq.~(\ref{eqn:Response}) and quality factor and resonance frequency are extracted. 
	To approximately bias the cavity with a desired value for $\partial\omega_0/\partial\Phi$, we run this biasing and fitting procedure until the resonance frequency shift between two subsequent bias points matches the set value.
	\textbf{III.} Then, we switch on the drive tone at a frequency $\omega_d$ slightly below the red sideband frequency $\omega_0 - \Omega_m$ with $\omega_0$ being the last resonance frequency measured in the bias flux sweep, and move the drive tone frequency in small steps towards the cavity resonance frequency.
	For each pump frequency, we take a resonance curve and extract $\omega_0$ by a fit again, until $\omega_0 - \omega_d - \Omega_m < \kappa/100$, i.e., until the drive tone is approximately on the red sideband.
	\textbf{IV.} When this criterion is fulfilled, the iteration stops, we switch off the pulse-tube cooler of the dilution refrigerator and measure one full cavity transmission spectrum as well as a narrow-band zoom-in transmission to the frequency range where the transparency occurs $\Omega \approx \Omega_m$.
	This relatively complicated iterative procedure is needed for several reasons.
	First, due to the non-negligible loop inductance and the possibly non-sinusoidal current-phase relation, we operate the cavity for most measurements in a metastable and hysteretic biasing regime. 
	Second, the cavity resonance frequency depends slightly on the intracavity photon number despite the small anharmonicity.
	Many parameters such as the flux sweetspot biasing value or the sweetspot frequency depend furthermore slightly on the in-plane field value, what we attribute mainly to an imperfect alignment between sample and magnetic in-plane field, leading to a non-negligible out-of-plane component.
	Taking all these factors together, a simple fixed biasing procedure to achieve similar parameters for each measurement would not be sufficient.

	\subsubsection{Fitting routine}

\begin{figure}
		\centerline {\includegraphics[trim={2cm 6cm 2cm 1cm},clip=True,width=0.6\textwidth]{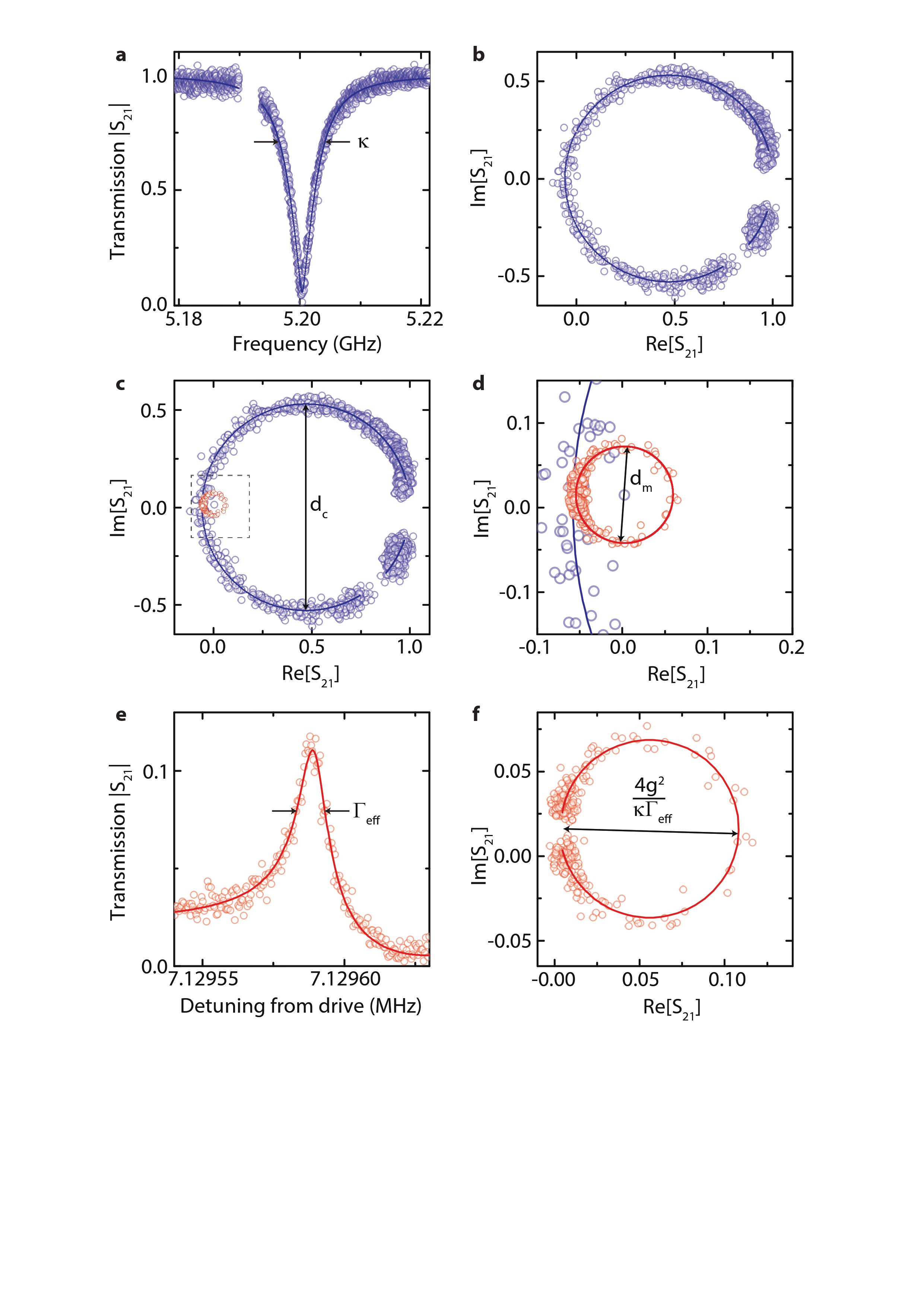}}
		\caption{\textsf{\textbf{Fitting the optomechanical response and extracting the multi-photon coupling rate $g=\sqrt{n_c}g_0$.} \textbf{a} Fit of the cavity response amplitude in presence of a red-sideband drive tone. The frequency window of the drive tone is removed for a reliable fitting procedure. \textbf{b} Dta a nd fit as in \textbf{a}, shown in the complex plane. \textbf{c} The large circle corresponds to the cavity response, the small circle to the signal of the optomechanically induced transparency, which is measured separately due to the narrow mechanical linewidth. The diameter of the cavity circle is $d_c$. The dashed box shows the zoom window plotted in \textbf{d}, where the diameter of the OMIT circle is denoted as $d_m$. In addition to a simple circle fit as represented by the line in \textbf{d}, we perform a fit of the complex resonance function to extract the effective mechanical linewidth. The result for the amplitude is shown in \textbf{e} and in the complex plane in \textbf{f}. Note that the data in \textbf{e} and \textbf{f} have been shifted and rescaled in the complex plane with respect to \textbf{c} and \textbf{d}. The scaling has been performed to anker the cavity circle at $S_{21} = 1$ with $d_c =1$. With this scaling the amplitude of the OMIT response is given by $4g^2/\kappa\Gamma_\mathrm{eff}$ as indicated in \textbf{f}.}}
		\label{fig:OMITFit}
	\end{figure}

	\textbf{I.} For the extraction of the single-photon coupling rate $g_0$ we initially perform a wide range scan as described in Sec.~\ref{sec:BGFitting} and get the background fit function $S_\mathrm{back}=P(\omega)e^{i\phi(\omega)}$.
	For all other measurements, we then calculate the complex background signal for the corresponding frequency range and divide it off the data.
	\textbf{II.} To fit the resonance curve for each measurement, the pump tone signal, which lies within the cavity line due to $\kappa \sim \Omega_m$, is cut away and the result is fitted as described in Sec.~\ref{sec:BGFitting} in order to obtain resonance frequency $\omega_0$ and linewdith $\kappa$.
	One example is shown in Fig.~\ref{fig:OMITFit}\textbf{a} and \textbf{b}.
	\textbf{III.} For the analysis of the transparency window, once again the backgrounds are divided off in a similar way as previously done for the cavity.
	During the cavity fit, the parameters $K$ and $\theta$ are determined and the cavity resonance was corrected for them, anchoring the resonance circle at $S_{21} = 1$.
	In addition, we apply all corrections to the mechanical response as well.
	An example for the real and imaginary part of both modified cavity and OMIT response functions are shown in Fig.~\ref{fig:OMITFit}\textbf{c} and \textbf{d}.
	Performing a circle fit as shown in Fig.~\ref{fig:OMITFit}\textbf{d}, we get the circle diameter $d_m$.
	\textbf{IV.} From a response fit to the mechanical resonance, we finally extract the last missing parameters $\Gamma_\mathrm{eff}$ and $\Omega_m$.
	At this stage, we can also determine the detuning between the cavity resonance frequency and the OMIT resonance $\delta_m$, which can be seen in Fig.~\ref{fig:OMITFit} as slight rotation of the OMIT response along the cavity circle and a Fano-like resonance in Fig.~\ref{fig:OMITFit}\textbf{e}.
	For the resonance shown in the main Fig.~3, we manually corrected for this rotation.
	\textbf{V.} Now we calculate the effective cooperativity and the multi-photon coupling rate $g$.
	The single-photon coupling rate $g_0$ is determined in the last step from $g$ using the independently calculated intracavity photon number $n_c$.

	\section{Data scaling, additional data and error bars}
	
	\subsection{Accounting for deviations in $\partial\omega_0/\partial\Phi$}

	As mentioned above, we face the complications that the used SQUID cavity is metastable and hysteretic, that the out-of-plane component of the in-plane field slightly influences the cavity parameters and its flux-dependence, that the sweetspot frequency is slightly varying with magnetic history and that the biasing current is (partly) flowing through the mechanical oscillator.
	To work around these effects, we automatically bias sweep the cavity prior to each measurement until the desired $\partial\omega_0/\partial\Phi$ is approximately achieved.
	In most cases, however, the real $\partial\omega_0/\partial\Phi$ slightly deviates from the set value due to the fitting error of $\omega_0$ and due to a non-constant conversion from bias current to flux in particular for higher in-plane fields $B_{||} \sim 10\,$mT.
	Typically, this error is around $10\%$ of the set value.
	To compensate for this, we additionally fit the flux dependence taken during the search for the desired $\partial\omega_0/\partial\Phi$ and extract a more precise number for the real value from there. 

	\subsection{Additional data set for $g_0$ vs $\partial\omega_0/\partial\Phi$}

\begin{figure}
		\centerline {\includegraphics[trim={2cm 3cm 2cm 4cm},clip=True,width=0.75\textwidth]{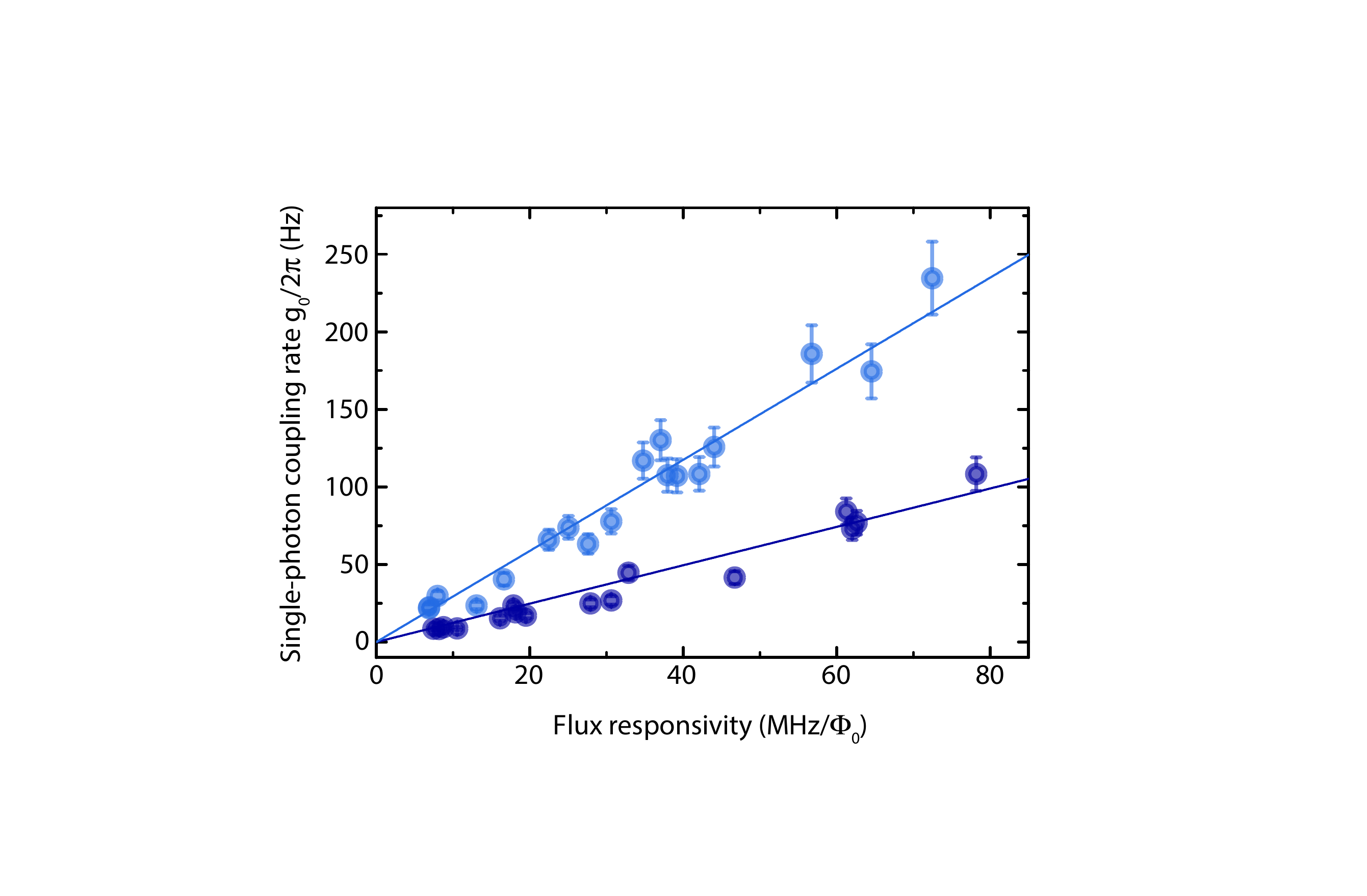}}
		\caption{\textsf{\textbf{Tuning of the single-photon coupling rate $g_0$ with flux responsivity.} The bright blue points correspond to the values for $g_0$ extracted for $B_{||} = 10\,$mT and the dark points to $B_{||} = 5\,$mT. The lines correspond to the theoretical calculations, where the line for $5\,$mT has been scaled by a factor of 0.83.}}
		\label{fig:VsdwdPhi}
	\end{figure}

	In Fig.~\ref{fig:VsdwdPhi} we show the data from the main paper Fig.~3\textbf{e} together with a second data set obtained for $B_{||} = 5\,$mT, here plotted vs $\partial\omega_0/\partial\Phi$ instead of flux bias to demonstrate the linear dependence more clearly.
	Both datasets show an approximately linear increase and the ratio between the slopes of the two theoretical lines $2.4$ is close to the expected value of $2$ arising from the difference in in-plane field.
	The deviation from the factor 2 seems to be related to a systematic influence of the in-plane field to the system as can also be seen in Fig.~4 of the main paper, where the datapoints for $5\,$mT lie under the theoretical line, while the points for $10\,$mT lie slightly above. 

	\subsection{Main paper error bars}

	The error bars given in the main paper on the data points in Fig.~3\textbf{e} and Fig.~4\textbf{c} are estimates of $10\%$ of the extracted value for $g_0$.
	The dominant factor for this uncertainty is given by possible variations of the feedline input power with frequency.
	Typically, we find cable resonances in our setup around $2$ to $3\,$dB, cf. Fig.~\ref{fig:Fitting}\textbf{a}.
	Due to our experimental scheme, where the red sideband drive tone is sent to the sample through another cable than the probe tone, cf. Fig.~\ref{fig:Setup}\textbf{a} and \textbf{d}, and the drive signal is not going to the amplifier chain, we cannot calibrate for possible cable resonances on the drive tone.
	We roughly calibrated the input attenuation of the drive line by disconnecting the sample and measuring the reflection of the open connector at base temperature into the amplifier line.
	The cable resonances, however, might significantly differ, when the sample is connected and thus the frequency dependence is not accessible. 
	This leaves us with possible small uncertainties regarding the intracavity photon number.
	Additional sources for errors lie in the fit values of the cavity linewidth, the OMIT amplitude and the mechanical linewidth.

	\subsection{Accounting for uncertainties in $\partial\omega_0/\partial\Phi$ in the in-plane dependence}

	In order to see the linear scaling of $g_0$ with the in-plane magnetic field, the data for different in-plane fields have to be taken for all other conditions fixed, in particular for the flux responsivity $\partial\omega_0/\partial\Phi$ being constant.
	This biasing procedure is non-trivial in our device due to the metastability of the SQUID, an out-of-plane component of the in-plane field and slightly field-dependent biasing conditions. 
	Therefore, we only give a range for the expected theoretical $g_0$ in the main paper Fig.~4\textbf{c}, assuming possible variations of $\partial\omega_0/\partial\Phi$ of about $10\%$.

\end{document}